\documentclass[letterpaper]{JHEP3}

\pdfoutput=1

\newcommand{\tr}{\mathrm{tr}}

\newcommand{\beq}{\begin{equation}}
\newcommand{\eeq}{\end{equation}}

\usepackage{graphicx}
\usepackage{amsfonts}
\usepackage{amsmath}
\usepackage{epsfig}

\title{Extracting black hole physics from the lattice}

\author{Simon Catterall\\
{\it Department of Physics, Syracuse University, Syracuse, NY13244, USA } \\
\email{smc@physics.syr.edu}
}
\author{Toby Wiseman\\
{\it Theoretical Physics Group, Blackett Laboratory, Imperial College, London SW7 2AZ, UK } \\
\email{t.wiseman@imperial.ac.uk}
}

\date{September 2009}

\abstract{
We perform lattice simulations of $N$ D0-branes at finite temperature in the decoupling limit, namely 16 supercharge $SU(N)$ Yang-Mills quantum mechanics in the 't Hooft limit. At low temperature this theory is conjectured to be dual to certain supergravity black holes. We emphasize that the existence
of a non-compact moduli space
renders the partition function of the
quantum mechanics theory divergent, and we perform one loop calculations that demonstrate this explicitly.
In consequence we use a scalar mass term to regulate this divergence and argue that the dual black hole thermodynamics may be recovered in the appropriate large $N$ limit as the regulator is removed. We report on simulations for $N$ up to 5 including the Pfaffian phase, and $N$ up to 12 in the phase quenched approximation. Interestingly, in the former
case, where we may calculate this potentially difficult phase, we find that it appears to play little role dynamically over the temperature range tested, which is certainly encouraging for future simulations of this theory.
}

\begin{document}

%
\section{Introduction}
%

There are currently various claimed perturbatively renormalizable completions of gravitational theories, including perturbative closed strings, $\mathcal{N} = 8$ supergravity \cite{Bern} and Horava's recent gravity proposal
\cite{Horava}. Whilst attaining such finite or renormalizable behaviour is a crucial step, given any such  theory one would further
like to explore the fascinating non-perturbative gravitational phenomena that are believed to exist from semiclassical reasoning. In the context of string theory, D-branes provide powerful insights into non-perturbative quantum gravity as in certain contexts these objects are thought to have two equivalent descriptions; that given by closed strings which lead to a gravitational theory, and that of open strings which yield a non-gravitational theory. Specifically Maldacena's duality conjectures that a large number $N$ of coincident D$p$-branes in the `decoupling' limit \cite{Mald,Itzhaki} is described by closed strings close to a near extremal charged black hole carrying $D$-brane charge and tension, and in the open string picture
by a $(p+1)$-dimensional strongly coupled
maximally supersymmetric $SU(N)$ Yang-Mills theory, taken in the 't Hooft limit. Whilst the closed string or gravitational description is problematic to quantize using perturbation theory, since the open description is non-gravitational, its quantization is, at least in principle, well understood using conventional field theory methods. 

Such dualities remain conjectural, although the weight of evidence supporting them is by now convincing. However relatively little concrete information has emerged about the non-perturbative nature of the dual quantum gravity, essentially because this appears to require solution of the strongly coupled field theory. Unfortunately there has been little progress analytically beyond the planar limit which cannot capture the interesting non-perturbative gravitational physics. 
The notable exception to this is the case of D1-D5 branes where the entropy may be deduced by computing an index which may be calculated by deforming the theory to weak coupling \cite{StromVafa, Strom}. However the use of indexes has unfortunately so far not been useful for computing non-perturbative phenomena in the context of D3 branes \cite{MaldMinw}.

Thus it appears that we may have to take numerical approaches or consider new analytic approximation schemes, such as the Gaussian approximation of Kabat, Lifschytz and Lowe \cite{Kabat1,Kabat2}, to solve these theories if we hope to directly extract information about quantum gravity. 
A natural context to start with is to consider the finite temperature theory, so that the dual gravity in appropriate limits describes the full quantum behaviour of black holes. A first step that is the subject of this paper is to extract the crudest information possible, the thermodynamics of the theory, which presumably should yield the dual Bekenstein-Hawking entropy. We will address this in the most amenable numerical context for the holographic correspondence, the case of D0 branes, and will attempt to simulate the thermal partition function using lattice methods.
Here the open string description is the 16 supercharge Yang-Mills quantum mechanics (the `BFSS model' \cite{BFSS}). This model is particularly
tractable for several reasons the most important of
which is that the theory is not only super-renormalizable, but is also finite in the continuum limit when regulated on the lattice. Thus the usual
problems of fine tuning can be avoided and there is no necessity
to utilize the more exotic supersymmetric lattice
constructions described in \cite{Catterall:2009it}.

Numerical simulation of this theory at finite temperature has been recently performed on the lattice \cite{Catterall:2008yz} and using a Fourier cut-off\cite{Nishimura16,Hanada:2008gy,
Hanada:2008ez,Kawahara:2007ib}.\footnote{See also related zero temperature numerical works \cite{Campostrini1, Hiller, Lunin} in this and related theories.} However these works ignored an important physical point, namely that the thermal partition function is actually divergent due to the exact quantum moduli space enjoyed by D0 branes, as first observed by Kabat, Lifschytz and Lowe \cite{Kabat1,Kabat2}. This divergence is associated with the regions of field space where the classical moduli become well separated, and depending on the temperature probed, such divergences may be hard to see in a Monte Carlo simulation. 

In this paper we use 1-loop methods to examine the divergence in the continuum theory and claim that it exists at for any $N$ and any temperature. We propose regulating the divergence with a mass for the scalar fields in the theory, and argue how to recover the relevant dual gravitational physics as the regulator is removed. We then perform lattice simulations with the mass regulator, and explicitly demonstrate this procedure. We pay particular attention to the potentially problematic Pfaffian phase -- the `sign problem' -- that arises in our simulations (and indeed in any 16-supercharge Euclidean Yang-Mills simulations), and find that in the thermal theory the phase fluctauations are fortunately rather small over the temperature range probed.

%
\section{Brief review of duality with black holes}
%

Following Itzhaki et al  \cite{Itzhaki}, we consider the ``decoupling'' limit of $N$ coincident D0-branes. We take $N$ large with $N g_s$ fixed, where $g_s$ is the string coupling. The decoupling limit is then defined by considering excitations of these D0-branes with fixed energy while sending the string length scale to zero so $\alpha' \rightarrow 0$. In this limit the degrees of freedom of the system split up into those localized near the branes - the `near horizon' excitations - and those living far from the brane which we are not interested in here.

There are two descriptions of the degrees of freedom living near the branes in the decoupling limit. The first, the open string description, arises from the open string worldvolume theory of the D0-branes whose degrees of freedom are the open strings ending on the branes. In the limit of fixed energy excitations as $ l_s \rightarrow 0$, the dynamics is governed by 16 supercharge $SU(N)$ Yang-Mills quantum mechanics with gauge coupling  $g_{YM}^2 = g_s \alpha'^{-3/2} / (2\pi)^2$.
Explicitly this theory arises from dimensional
reduction of $\mathcal{N} = 1$ super Yang-Mills in 10-d. The
10-d gauge field reduces to the 1-d gauge field $A$ and 9 scalars,
$X^i$, $i = 1,\ldots,9$ and the 10-d Majorana-Weyl fermion to
16 single component fermions, ${\Psi}_{\alpha}$, $\alpha=1,\ldots,16
$. The action is given as,
\begin{eqnarray}
S & = & \frac{N}{\lambda} \tr \int dt
        \left\{ \frac{1}{2} (D_{t} X_i)^2 - \frac{1}{4} \left[ X_i,
X_j \right]^2 + \Psi \gamma^t \left( \gamma^t D_{t}  - \gamma^i \left[ X_i, \cdot
\right] \right) \Psi \right\}  
\end{eqnarray}
where $\gamma^t, \gamma^i$ are the real Majorana-Weyl
gamma matrices.

The second description is the closed string one, namely IIA closed strings propagating in the near horizon geometry of $N$ D0-branes. When supergravity is valid we may write the vacuum near horizon geometry as,
\begin{eqnarray}
ds^2 & = & \alpha' \left( \frac{U^{\frac{7}{2}}}{2 \pi \sqrt{b \lambda}} (-dt^2) + 2 \pi \sqrt{b \lambda} \left( U^{-\frac{7}{2}} {dU^2} + U^{-\frac{3}{2}} d\Omega^2\right) \right) \nonumber \\
e^\phi & = & (2 \pi)^2 g_{YM}^2 \left( \frac{U^7}{4 \pi^2 b \lambda} \right)^{-\frac{3}{4}}
\end{eqnarray}
where $b = 240 \pi^5$, with $\lambda = N g_{YM}^2$ and the coordinate $U$ can now be interpreted as an energy scale. Closed string excitations may be added to this and then the geometry is simply asymptotic (as $U \rightarrow \infty$) to this. 
It is crucial that all curvatures and the string coupling $e^\phi$ are small in order that the above supergravity solution is valid. The curvature radius $\rho$ at energy scale $U$ is characterized by the radius of the sphere in the above geometry, so that in string units,
\begin{equation}
\frac{\rho}{\alpha'^{1/2}} \sim \left( \frac{\lambda}{U^3} \right)^{1/4}
\end{equation}
and the dilaton at the radius $U$ is,
\begin{equation}
e^{\phi} \sim  \frac{1}{N} \left(\frac{\lambda}{U^3} \right)^{\frac{7}{4}}
\end{equation}
Hence we see that provided $\lambda/U^3$ and $N$ are large the supergravity solution above is a good approximation. In particular the dilaton condition shows we must take the 't Hooft limit, first taking $N$ to infinity with $\lambda/U^3$ fixed, and subsequently taking this ratio large. Moving to energies
where $U^3/\lambda \sim 1$, curvature corrections become important. At ultra low energies, $U^3/\lambda \sim N^{-4/7}$, outside the 't Hooft scaling limit, the dilaton blows up and the string theory becomes strongly coupled. In
conclusion the closed string description in vacuum reduces to the above
supergravity solution above in the approximate range $N^{-4/7} < U^3 / \lambda < 1$.

At finite temperature the relevant near horizon geometry in the supergravity approximation is a black hole,
\begin{eqnarray}
ds^2 & = & \alpha' \left( \frac{U^{\frac{7}{2}}}{2 \pi \sqrt{b \lambda}} (-f dt^2) + 2 \pi \sqrt{b \lambda} \left( U^{-\frac{7}{2}} \frac{dU^2}{f} + U^{-\frac{3}{2}} d\Omega^2\right) \right) \nonumber \\
f(U) & = & 1 - \frac{U_0}{U}
\end{eqnarray}
and the horizon is located at $U = U_0$, with $U_0$ parametrically giving the thermal energy scale.
In this decoupling limit the energy above extremality, E, and temperature T become,
\begin{equation}\label{eq:YMerg}
\epsilon  = E/\lambda^{1/3} = N^2 \frac{8}{7 (2 \pi) b} \left( U_0 / \lambda^{1/3} \right)^7 , \qquad t = T/\lambda^{1/3} = \frac{7}{8 \pi^2 \sqrt{b}} \left( U_0 / \lambda^{1/3} \right)^{5/2} 
\end{equation}
where we have defined the natural dimensionless energy and temperature variables $\epsilon$ and $t$. Again we can only trust this solution where string loop and curvature corrections are small. Choosing the thermal scale to be $U_0^3 / \lambda \sim O(1)$ so $t$ remains non-zero and finite as we take $N$ to infinity hides the region of the geometry where the dilaton becomes large behind the horizon. For small values of $U_0^3/ \lambda << 1$, so $t << 1$,
then the range of the geometry $U_0^3 / \lambda \le U^3 / \lambda < 1$  is well described by supergravity while for larger $U$ curvature corrections contribute. We may then reliably  compute the entropy as a function of temperature since the entropy depends only on the near horizon geometry. This yields a prediction for the thermodynamic behaviour,
\begin{eqnarray}
\epsilon = c\; N^2 t^{14/5} \qquad c = \left( \frac{2^{21} 3^{12} 5^
{2}}{7^{19}} \pi^{14} \right)^{1/5} \simeq 7.41 .
\end{eqnarray}
valid for $t << 1$ but finite in the large $N$ limit. Interestingly the recent work of Smilga \cite{Smilga:2008bt} may provide an argument for the origin of this  power law dependence on $t$. 
If we move to higher temperatures we reach the Horowitz-Polchinski correspondence point \cite{HP} at $t \sim 1$ where the supergravity breaks down even at the horizon due to curvature corrections. Note however that in principle a closed string description still exists, although we have little control over it.

For ultra low temperatures outside the 't Hooft scaling limit, $t << 1/N^{-10/21}$, the dilaton becomes large near the horizon so that the supergravity description breaks down. However, the strongly coupled IIA theory may be lifted to M-theory where one finds 11-d supergravity is valid and thermodynamic predictions can be made \cite{Itzhaki}. However we will not be concerned with this M-theory limit here.

%
\section{Divergence of the thermal partition function}
%

The results of Paban, Sethi and Stern indicate that the Lorentzian 16 supercharge quantum mechanics possesses an exact quantum moduli space \cite{Sethi}, given by the space of commuting scalar and fermion fields. These may always be simultaneously diagonalized by a gauge transformation and then the bosonic moduli space is parameterized by the $N-1$ diagonal entries of the 9 scalars, which can be thought of as giving the moduli space of $N-1$ identical particles which describes the D0-branes minus their centre of mass motion. Paban, Sethi and Stern constructed the effective action for one of these particles when it is far separated from the others and showed that supersymmetry constrained the form of this action to have vanishing potential when the particle is stationary, which is maintained under adding corrections due to the other particle interactions. Another reflection of this result is the work of de Wit, L\"uscher and Nicolai  \cite{deWit} who showed that the theory possesses a continuum of states extending down to zero energy. 

Given that the quantum mechanics has an exact \emph{non-compact} quantum moduli space and a corresponding continuum of states extending to zero energy, it follows that the thermal partition function for $N$ D0-branes is likely ill-defined at low temperatures due to the infra-red divergence associated to the integral over this moduli space. As far as we are aware, Kabat, Lifshitz and Lowe \cite{Kabat1,Kabat2} provide the first statement in the literature that the thermodynamic partition function suffers from such a divergence.
An important point is that while the thermal path integral is divergent, the Euclidean path integral with periodic boundary conditions for the fermions, should be well defined, giving the Witten index. Since one might naively imagine that the low temperature thermal behaviour would resemble the periodic Euclidean theory for large Euclidean time radius, we see that the existence of a divergence in the thermal case, but not in the periodic case, implies the divergence is a somewhat subtle effect.

At low temperature, the continuous spectrum down to zero energy might suggest a divergence -- although of course it is possible to have a continuum of states and not have a divergence. However, it is less clear that this persists to higher temperatures. Kabat, Lifshitz and Lowe stated such a divergence exists in the $t \sim O(1)$ temperature range although they did not consider it in detail as the Gaussian approximation they employed elegantly sidesteps the issue. We are unaware of any detailed discussion in the literature concerning this divergence, which seems somewhat surprising given its importance
for the thermodynamics of this maximally supersymmetric YM quantum mechanics,  or indeed the Yang-Mills theory associated with Dp-branes on a 
torus which one assumes would suffer the same instabilities for the same reasons.

The physics of the divergence is presumably understood in terms of Hawking evaporation of D0-branes from the black hole, since it is clearly associated with the non-compact moduli space of the branes.\footnote{We note that our interpretation differs from that of \cite{Kabat1,Kabat2} where the effect is stated to be dual to thermal gravitons.} We have little to say about this closed string interpretation, although it certainly appears interesting, and it would be intriguing to see if it could be calculated in the closed string picture, although as it involves thermal radiation of D0-branes, it could presumably not be seen at the level of supergravity itself. 

The aim of the rest of this section is to use the 1-loop approximation to study the thermal dynamics of the model and show that an IR divergence should indeed be expected to arise. 
A simple 1-loop argument was given by Anagnostopoulos et al \cite{Nishimura16} that nicely illustrates  the nature of the divergence. We extend this argument by performing a detailed 1-loop calculation for the bosonic version of the theory, and in particular compute the first non-trivial quantum contributions to the classical moduli space from \emph{all} fluctuations about the moduli (rather than just the `off-diagonal' fluctuations as in previous 1-loop calculations \cite{Ambjorn:1998zt} used in the argument of \cite{Nishimura16}). We use the structure of this bosonic result to argue that in the supersymmetric case a divergence exists.

Following the 1-loop argument in \cite{Nishimura16} of an infrared instability, it was claimed that in the Monte Carlo simulation the instability was seen at low temperature but eliminated by increasing $N$. Here we wish to emphasize that the instability is intrinsic to the continuum theory at all temperature and $N$, and must be seen in the Monte Carlo (if one can run the simulation long enough). It is therefore important to regulate this divergence on the lattice carefully.

%
\subsection{The matrix integral truncation: classical moduli and the 1-loop approximation}
\label{sec:matint}
%

In this section we will outline the nature of the 1-loop
approximation in the context of a simpler model namely bosonic
Yang-Mills matrix theory. This model arises
by compactification of the thermal quantum mechanics
model to zero dimensions.
In the bosonic sector the starting Euclidean action is simply given as,
\begin{eqnarray}
S & = & \frac{N}{\lambda} \tr \oint^R d\tau
        \left\{ \frac{1}{2} (D_{\tau} X_i)^2 - \frac{1}{4} \left[ X_i,
X_j \right]^2  \right\}
\end{eqnarray}
where $R$ is the length of the periodic Euclidean time $\tau$, and gives the temperature $T$ as $T = 1/R$.
Taking
the fields to be constant in time, $A(\tau) =  \tilde{X}_0, X_i(\tau) = \tilde{X}_i$ yields the Yang-Mills matrix integral,
\begin{eqnarray}
S & = & -\frac{N R}{4 \lambda} \tr \left\{
         \left[ \tilde{X}_\mu, \tilde{X}_\nu \right]^2  \right\}  
\end{eqnarray}
which having no coupling constant 
(the factor $R/\lambda$ appearing can simply be absorbed by a rescaling of the fields) is obviously strongly coupled.
This matrix integral has a classical moduli space of commuting matrices, which lift to the classical moduli of our full quantum mechanics which are given by commuting bosonic matrix fields which are constant in time. Whilst the full dynamics of this matrix integral is strong coupled there is a regime where the dynamics is weakly coupled and may be studied by using the 1-loop technique \cite{Aoki:1998vn,Hotta:1998en,Krauth:1999qw}.

By a choice of gauge we may choose the vacuum configuration to be
given by a set of diagonal matrices. 
We decompose a Hermitian matrix into its diagonal and off-diagonal parts, using the notation $Z =  z + \hat{Z}$ where $z$ is diagonal and $\hat{Z}$ off diagonal. We expand the matrices about such a diagonal configuration,
\begin{eqnarray}
\tilde{X}_{\mu} & = & x_\mu  +  \hat{X}_{\mu} 
\end{eqnarray}
where the classical bosonic moduli are the diagonal components of $x_\mu$ which we denote $x^a_\mu$. A classical solution is then some choice of $x_\mu$ and zero off-diagonal modes $\hat{X}_\mu = 0$. Then with the notation that $\Delta x^{ab}_\mu  = x^a_\mu - x^b_\mu$ and $|\Delta x^{ab}|^2 = \sum_\mu (\Delta x^{ab}_\mu)^2$, the action becomes,
\begin{eqnarray}
S =  && \frac{N R}{\lambda} \left( \sum_{a < b} (\hat{X}_\mu)^{*}_{ab} \left( \delta_{\mu\nu}  | \Delta x^{ab}|^2 - \Delta x^{ab}_\mu \Delta x^{ab}_\nu \right) (\hat{X}_\nu)_{ab}  \right. \nonumber \\
      && \hspace{5cm} \left.   - \tr \left(  [ x_\mu , \hat{X}_\nu ] [ \hat{X}_\mu , \hat{X}_\nu ] + \frac{1}{4} [ \hat{X}_\mu , \hat{X}_\nu ]^2 \right) \right)
\end{eqnarray}
where we $(\hat{X}_\mu)_{ab} = (\hat{X}_\mu)^*_{ba}$ as they are Hermitian. 

We may think of the off-diagonal matrix $\hat{X}_\mu$ as being composed of the degrees of freedom of the $N(N-1)/2$ complex scalars $(\hat{X}_\mu)_{ab}$ with $a < b$, with the first term above giving their quadratic action, while
the latter two terms give their interactions. We see that the off-diagonal modes, $(\hat{X}_\mu)_{ab}$, gain a mass $| \Delta x^{ab} |$ from the classical moduli.\footnote{We see from the Greek index structure that there also is a zero mass fluctution of the $\hat{X}_\mu$ due to the gauge not being fixed here which we treat in detail in Appendix A.} If this mass is very large, we may use the 1-loop approximation to the system by taking the diagonal fields as slow degrees of freedom, and the off-diagonal fields as fast degrees of freedom. 
Let us rescale the fields as follows;
\begin{eqnarray}
x^a_\mu & = & {\Lambda} \phi^a_\mu \nonumber \\
( \hat{X}_\mu )_{ab} & = & \frac{1}{\Lambda} \sqrt{\frac{\lambda}{R}} ( \hat{\Phi}_\mu )_{ab} \; .
\end{eqnarray}
where $\Lambda$ is a mass scale.
Now the classical moduli fields $\phi^a_\mu$ and the off diagonal modes $(\hat{\Phi}_\mu)_{ab}$ are dimensionless.
Consider giving values to the classical moduli so that $\phi^a_\mu \sim O(1)$. Then the mass scale $\Lambda$ sets the mass for the off-diagonal modes. We find their action to be,
\begin{eqnarray}
S &=&  N \left( \sum_{a < b} (\hat{\Phi}_\mu)^{*}_{ab} \left( \delta_{\mu\nu} 
 | \Delta \phi^{ab}|^2 - \Delta \phi^{ab}_\mu \Delta \phi^{ab}_\nu \right) (\hat{\Phi}_\nu)_{ab} \right. \\
&-&\left. \tr \left(  g [ \phi_\mu , \hat{\Phi}_\nu ] [ \hat{\Phi}_\mu , \hat{\Phi}_\nu ] + \frac{1}{4} g^2 [ \hat{\Phi}_\mu , \hat{\Phi}_\nu ]^2 \right) \right)
\end{eqnarray}
with dimensionless coupling, $g = \lambda / (R \Lambda^4)$. Thus for a given coupling $\lambda$ and temperature $1/R$, provided we choose the mass scale $\Lambda$ to be large enough,
\begin{eqnarray}
\Lambda >> \left( \frac{\lambda}{R} \right)^{1/4}
\end{eqnarray}
so that $g << 1$, we may integrate out the off diagonal modes in the 1-loop approximation.

This yields an effective theory for the classical moduli $\phi^a_\mu = \frac{1}{\Lambda} x^a_\mu$. Whereas classically there is no potential for these  moduli fields,
in the case of this matrix integral, these massive off diagonal modes generate a 1-loop potential. 
As we have seen, we may trust this 1-loop potential when the moduli have values, $| x^a_\mu | >> \left( \frac{\lambda}{R} \right)^{1/4}$ so that the mode separations,
\begin{eqnarray}
R | \Delta x^{ab} | >> \left( \lambda R^3 \right)^{1/4}
\end{eqnarray}
and therefore the masses generated for the off diagonal modes, which go as $| \Delta x^{ab} |^2$, are sufficiently large. Thus it is in the region where the 
moduli are well separated that the 1-loop potential is a
good approximation.

The classical moduli space has, in an appropriate gauge, a measure $\int \prod_{a = 1}^{N-1} \prod_{\mu = 0}^{9} dx^a_\mu$ where we note that $x^N_\mu$ is determined by the traceless condition on the original matrices $X_\mu$ due to the $SU(N)$ gauge group. We see the classical moduli have infinite volume in their measure.
Without a potential, this would lead to a divergence in the matrix integral from the region where these modes are well separated. However, it is precisely here that one can trust the 1-loop calculation above, and one finds the 1-loop potential is attractive, (see eg. \cite{Aharony2})
\begin{eqnarray}
S_{1-loop} \sim \sum_{a<b} \log (R | \Delta x^{ab} |)
\end{eqnarray}
and therefore renders the integral over these classical moduli finite. We note that for supersymmetric matrix integrals, 1-loop calculations again suggest convergence \cite{Aoki:1998vn,Hotta:1998en,Krauth:1999qw}, and in these cases remarkable analytic work has proven the integral to be finite \cite{Austing:2001pk, Austing}.

Thus we see that whilst naively the theory is always strongly coupled, for large diagonal separations $| \Delta x^{ab} | >>  \left( \frac{\lambda}{R} \right)^{1/4}$ of the matrix fields, the system may then be approximated by the dynamics of these diagonal modes with corrections from integrating out massive off diagonal modes.

%
\subsection{1-loop approximation for the quantum mechanics}
\label{sec:BO}
%

The 1-loop technique was pioneered by Douglas et al \cite{Douglas:1996yp} for the supersymmetric quantum mechanics, and was first used for finite temperature calculation by Ambjorn et al \cite{Ambjorn:1998zt} (see also \cite{Aharony2} for generalization to bosonic and higher dimensional theories).
Here we will show how to perform the 1-loop approximation for the bosonic quantum mechanics. We begin by expanding the matrix fields as,
\begin{eqnarray}
A(\tau) & = & a  \nonumber \\
X_i(\tau) & = & x_i + y_i(\tau) + \hat{X}_i(\tau)
\end{eqnarray}
where $a, x_i, y_i(\tau)$ are diagonal matrices, and $\hat{X}_i$ are off-diagonal matrices. We also make the choice that $\oint d\tau y_i(\tau) = 0$, and $a, x_i$ are constant in time. Note that we have made a choice of gauge so that $A(\tau)$ is diagonal and constant in time, which completely fixes the gauge freedom.

We now think of the decomposition as follows. The constant diagonal modes $a, x_i$ are the classical moduli. Setting the remaining fields to zero, the action for these moduli vanishes. As above, the off-diagonal and time dependent fields $\hat{X}_i$ will again be massive if we give values to the classical moduli, and these masses are given in terms of the classical moduli separation $| \Delta x^{ab}  |^2$, where now $|\Delta x^{ab}|^2 = \sum_i (\Delta x^{ab}_i)^2$. 
The new ingredient is the diagonal, non-constant modes $y_i(\tau)$. These modes are interesting as setting the off-diagonal fields to zero, their action is only quadratic and hence they are not coupled classically to the classical moduli. 

The procedure we follow is to firstly give large separations to the classical moduli, and integrate out the off-diagonal modes at 1-loop, as for the matrix integral above. This generates a 1-loop potential for the classical moduli $a, x_i$. However, it also generates  interaction terms for the diagonal, non-constant modes $y_i$ coupling them to the classical moduli. Since the only interaction terms for $y_i$ are generated from integrating out the massive off diagonal fields, these interactions are strongly suppressed relative to the classical kinetic terms for $y_i$. Thus we may now integrate out the diagonal non-constant modes $y_i$ also in the 1-loop approximation. The only interaction terms kept are then those that are quadratic in the $y_i$. These then yield a second contribution to the potential for the classical moduli. In previous literature (eg. \cite{Ambjorn:1998zt}) only the potential contribution from the off-diagonal modes has been computed, and since we are interested in convergence of the partition function it is important to compute the leading behaviour of potentials generated from \emph{all} fluctuations about the moduli as we do here.

We have saved the details of the calculation for Appendix A, and here present the main results. The action in our gauge is given as,
\begin{eqnarray}
S & = & \frac{N}{\lambda}  \oint^R d\tau \,
       \left(  \frac{1}{2} {(\partial_\tau y^a_i)}^2 + \frac{1}{2} ( \hat{X}^*_i )_{ab}  O^{ab}_{ij} ( \hat{X}_j )_{ab} + O( \hat{X}^3 ) \right)
\end{eqnarray}
where,
\begin{eqnarray}
O^{ab}_{ij} & =& \left( \hat{D}^{ab}_i \hat{D}^{ab}_j - \delta_{ij} (( \hat{D}^{ab}_\tau) ^2 + (\hat{D}^{ab}_k)^2 )\right) \nonumber \\
\hat{D}^{ab}_\tau &  =&  \partial_\tau + i \Delta a^{ab} \; , \qquad  \hat{D}^{ab}_i  = i ( \Delta x^{ab} _i + \Delta y^{ab}_i(\tau) )
\end{eqnarray}
and we have ignored the Fadeev-Popov determinant from the gauge fixing, which is treated in the Appendix. We see that as before, the classical moduli $x^a_i$ generate masses of order $| \Delta x^{ab} |^2$ for the off diagonal modes $(\hat{X}_i)_{ab}$, and the 1-loop approximation is again valid providing,
\begin{eqnarray}
R | \Delta x^{ab} | >> \left( \lambda R^3  \right)^{1/4}
\label{loopcond1}
\end{eqnarray}
as for the matrix integral above. Performing the 1-loop integration we find the action,
\begin{eqnarray}
S_{1-loop} & = & \frac{N}{2 \lambda}  \oint^R d\tau {(\partial_\tau y^a_i)}^2 + 4 \sum_{a \ne b} \log {\det} \left( \mathcal{O}^{ab}  + \epsilon^{ab} \right)
\end{eqnarray}
where,
\begin{eqnarray}
\mathcal{O}^{ab} & = & - ( \partial_\tau + i \Delta a^{ab} )^2 + | \Delta x^{ab} |^2  \nonumber \\
{\epsilon}^{ab} & = &  2 \Delta x^{ab}_i \Delta y^{ab}_i(\tau) + | \Delta y^{ab}(\tau) |^2
\end{eqnarray}
where the determinant is computed over the time circle $\tau$. 
This is exactly the analog of the result obtained for the supersymmetric theory in \cite{Ambjorn:1998zt}.
Since the masses $| \Delta x^{ab} |^2$ are large, we may further expand out this determinant, treating the operator $\epsilon^{ab}$ as a perturbation, giving, 
\begin{eqnarray}
S_{1-loop} & = & \frac{N}{2 \lambda}  \oint^R d\tau {(\partial_\tau y^a_i)}^2 + 4 \sum_{a \ne b} \log {\det} \mathcal{O}^{ab} \nonumber \\
&& \qquad \qquad  + 4 \sum_{a \ne b} {\tr} \left( (\mathcal{O}^{ab})^{-1}  {\epsilon}^{ab} - \frac{1}{2} (\mathcal{O}^{ab})^{-1} {\epsilon}^{ab} (\mathcal{O}^{ab})^{-1} {\epsilon}^{ab} + \ldots  \right) 
\end{eqnarray}
and we then decompose into Fourier modes,
\begin{eqnarray}
y_i(\tau) = \sum_{m=-\infty, m\ne0}^{\infty} y_{i(m)} e^{i (2 \pi / R)  m \tau }
\end{eqnarray}
in order to evaluate the determinants. We find,
\begin{eqnarray}
S_{1-loop} & = & V_0[ a, x_i] - \frac{2 \pi N}{R \lambda} \sum_{m = - \infty, m \neq 0}^\infty m^2 | y^a_{i(m)} |^2 \nonumber \\
&&   -  \sum_{a \ne b} \left( \sum_{m = - \infty, m \neq 0}^\infty \frac{R}{| \Delta x^{ab} |} \left( (\Delta y^{ab}_{i(m)})^* \left( \delta_{ij} - \frac{ R^2 \Delta x^{ab}_i \Delta x^{ab}_j }{ R^2 | \Delta x^{ab} |^2 + m^2 \pi^2 } \right) \Delta y^{ab}_{j(m)} \right) \right. \nonumber \\
&& \qquad \qquad \qquad  \left. + O\left[  \frac{R^2 ( | \Delta y |^2)^* | \Delta y |^2}{| \Delta x |^2} \right]  \right)
\end{eqnarray}
where we have defined $V_0[ a, x_i ] \equiv \sum_{a \ne b} \log {\det}_\tau \mathcal{O}^{ab}$ and are
keeping only quadratic terms in $y^a_i$. The expansion parameter that controls the size of the quantum correction terms to this action for $y^a_i$  relative to the quadratic classical kinetic term is $\lambda  R^2/ | \Delta x^{ab} |$. 
Thus we may integrate out the fields $y^a_{i(m)}$ at 1-loop, ignoring terms that are not quadratic in the $y^a_{i(m)}$ provided that,
\begin{eqnarray}
R | \Delta x^{ab} | >>   ( \lambda  R^3 )^{1/2}
\label{loopcond2}
\end{eqnarray}
We emphasize that we have included the quantum corrections to the classical quadratic action of $y^a_{i(m)}$ -- therefore going beyond the calculation in \cite{Ambjorn:1998zt} for the supersymmetric case -- 
as these include non-trivial interactions with the classical moduli (arising from the previous integration over the off diagonal modes). These then represent the leading interactions between the diagonal non-constant modes and the classical moduli in the limit of large moduli separation.
Finally performing the quadratic integral over $y_i$, we obtain,
\begin{eqnarray}
S_{1-loop} & = & V_0[ a, x_i ] + V_1[ a, x_i ]
\end{eqnarray}
where the first potential is derived from integration over the off-diagonal modes, and the second, from the subsequent integral over the diagonal non-constant modes. The conditions \eqref{loopcond1} and \eqref{loopcond2} for the two 1-loop integrals require,
\begin{eqnarray}
R | \Delta x^{ab} | >>  \max(  ( \lambda R^3)^{1/2} ,  \left( \lambda R^3 \right)^{1/4} )
\end{eqnarray}
which we may always satisfy for any $R, \lambda$ by simply taking large enough classical moduli separation.
When the dimensionless temperature $t = 1/( R \lambda^{1/3} )$ is small our approximation is valid for $R | \Delta x^{ab}  | >> 1 / t^{1/2}$ and at high temperature we require, $R | \Delta x^{ab}  |  >> 1 / t^{1/4}$.

For simplicity, we will give the behaviour of the potentials $V_0, V_1$ only for separations $R | \Delta x^{ab}  | >> 1$ as that is all we require here to consider convergence of the partition function. We note however that at high dimensionless temperature, actually our 1-loop approximation is good for $R | \Delta x^{ab}  | < 1$ and thus in principle we might have given $V_0, V_1$ for any separation. In this limit $R | \Delta x^{ab}  | >> 1$, as shown in Appendix A, the potentials are given as,
\begin{eqnarray} 
V_0[ a, x_i ] & \simeq & 8  \sum_{a < b} (R | \Delta x^{ab} |) + \ldots  \nonumber \\
V_1[ a, x_i ] & \simeq & \frac{32}{3} \lambda R^3 \sum_{a < b} \frac{1}{(R | \Delta x^{ab} |)} + \ldots 
\label{eq:oneloop}
\end{eqnarray}
where in the latter case
we have taken care to ensure that $\sum_a y^a_{i(m)} = 0$ when integrating over the fields $y^a_{i(m)}$ to enforce the traceless condition on the matrix fields $X_i$ due to the gauge group being $SU(N)$.

Now consider adding fermions to the theory. With periodic boundary conditions for the fermions, we have both bosonic and fermionic classical moduli, being the constant diagonal boson and fermion degrees of freedom. At 1-loop the
integration over off diagonal modes and non constant modes yields
equal and opposite contributions from boson and fermion sectors and the
effective potential vanishes.
However for thermal boundary conditions, there are no fermion zero modes, and the bosonic and fermion fluctuations about the bosonic classical moduli will not exactly cancel (due to the anti periodicity of the fermions). Thus we expect again a 1-loop action,
\begin{eqnarray}
S^{susy}_{1-loop} & = & V'_0[ a, x_i ] + V'_1[ a, x_i ]
\label{eq:BornOpp}
\end{eqnarray}
where $V'_0$ arises directly from integration over the bosonic and fermionic off diagonal modes, and $V'_1$ from subsequent integration over the diagonal non-constant fluctuations.

As we have seen above, already in the bosonic theory the contribution from the diagonal non-constant modes vanishes at large moduli separations. Thus in the supersymmetric theory, where we expect $| V'_1 | < | V_1 |$ due to boson/fermion cancellations we again will have a vanishing contribution at large separation. The important part of the potential is therefore $V'_0$, which can be explicitly evaluated giving,
\begin{eqnarray} 
V'_0[ a, x_i ] & = &  - 32 \sum_{a < b} e^{- R | \Delta x^{ab} |} \cos{ ( R \Delta a^{ab} ) }
\end{eqnarray}
in the large separation limit, $R | \Delta x^{ab} | >> 1$. Thus both $V'_0$ and $V'_1$ vanish at large classical moduli separation, the former exponentially, and the latter at least as fast as an inverse power in the separation. This means that the integration over the classical moduli space will yield a divergence arising from the region of the integral 
associated with widely separated moduli.

That $V'_0$ vanishes in the large moduli separation limit was the nice observation of \cite{Nishimura16} that led them to propose this gives rise to an instability. However, it is important to understand the nature of the leading corrections $V'_1$, as we have done here, as these could have given rise to a potential that could dominate $V'_0$ and lead to convergence.

It is an interesting fact that with no fermions the partition function is convergent and for periodic boundary conditions the partition function is also convergent (giving the Witten index), while for thermal boundary conditions it appears not to be. In the bosonic case as we have seen the off-diagonal modes provide a strong potential $V_0$ for the classical moduli which goes linearly in their separation. For periodic fermions while the potential arising from the massive modes cancel exactly due to supersymmetry, there are now fermion zero modes which counteract the potential divergence arising from the bosonic zero mode flat directions.\cite{Aoki:1998vn, four}

Since we might expect the thermal theory to resemble the periodic theory at low dimensionless temperature $t$, 
one might be concerned that the divergence we see from our thermal calculation is at odds with the fact that the periodic theory is convergent. However there is no disagreement when we recall that our calculation is only valid for separations $R | \Delta x^{ab}  | >> 1 / t^{1/2}$ for low temperature. Thus, our calculation must break down in the $t \rightarrow 0$ limit, as the region of field space where our approximation is valid is pushed out to infinite moduli separation. This break down is precisely due to the strong coupling of the diagonal non-constant modes, which in the supersymmetric case include the fermionic modes that, as $t \rightarrow 0$, look increasingly like fermionic zero modes.

Naive expectation also suggests that at high temperatures, $t >> 1$, the fermions in the thermal theory will have a thermal mass and the theory should resemble the bosonic theory, which is perfectly convergent.
However, we have seen this naive expectation is incorrect, and that
for any finite $t$ there is always a region of field space where the 1-loop approximation holds and where the potential on the moduli space vanishes asymptotically in separation, fast enough to yield a divergence.

We now summarize the results of this section. The exact quantum moduli space for the Lorentzian theory strongly suggests that the finite temperature theory should be IR divergent, at least at low temperature. We have used 1-loop methods to argue that this divergence should be present for \emph{all} finite temperatures.

%
\subsection{Regulating the thermal divergence}
%

As we have seen, we expect the SQM theory to have a thermal divergence associated with the regions of the classical moduli space of the Euclidean theory where the moduli are well separated. Schematically if we cut off the classical moduli space so that $R |\Delta x^{ab}| < r$ for some dimensionless regulator $r$ then from the above we expect the leading large $N$ behaviour of the Euclidean 
free energy to take the form,
\begin{eqnarray}
I \sim 9 N \log{r} + N^2 I_{finite}(\beta, r) .
\label{eq:divaction}
\end{eqnarray}
where upon removing the regulator we see a divergent term from the $9 (N-1)$ integrals over the non-compact classical moduli $x_i^a$. The leading $N^2$ part of the action should yield the physics dual to the black hole at low temperatures. Since it is this finite part of the action going as $N^2$ that we are really interested in, once the theory is regulated we may in principle extract the contribution $I_{finite}$ simply by taking $N$ large enough. However, we wish to obtain $I_{finite}$ in the limit of removing the regulator. An important point to note is that as we remove the regulator, the size of the regulated divergent term becomes larger, and hence in practice we require larger $N$ in order to extract the leading $N^2$ finite contribution of interest.
 
In the previous numerical work on this model \cite{Nishimura16,Catterall:2008yz,Hanada:2008ez,Hanada:2008gy} this divergence was ignored. In \cite{Nishimura16} the thermal instability of the simulation was said to be eliminated at low temperature by increasing $N$, and in \cite{Catterall:2008yz} the divergence was claimed to be a lattice artefact. This was consistent with the numerical data presented
in these papers because in practice it may be difficult to see this divergence, particularly when simulating at large $N$ (where the divergence is harder to see, and the simulations harder to run). As we have seen, the region of field space where the theory is strongly coupled is precisely the opposite region to that which contributes to the divergence, the weakly coupled region where the classical moduli are well separated. Hence if the lattice Monte Carlo algorithm does not sample this weakly coupled region
efficiently, one may deduce a possibly accurate estimation for the finite $O(N^2)$ contribution when $N$ is large. This presumably explains why 
this previous work found reasonable thermodynamic curves that look consistent with the predicted black hole thermodynamics at low temperature. However as we see later, if one runs the lattice simulation long enough then one must see the divergence. Hence our aim here is to regulate it carefully.
 
To regulate the theory we introduce a mass for the 9 scalars. We therefore take the Euclidean action,
\begin{eqnarray}
S & = & \frac{N}{\lambda} \mathrm{Tr} \oint^R d\tau
        \left\{ \frac{1}{2} (D_{\tau} X_i)^2 - \frac{1}{4} \left[ X_i,
X_j \right]^2 + \frac{1}{2} \mu^2 X_i^2 + \Psi \gamma^ \tau \left( \gamma^t D_{\tau}  - \gamma^i \left[ X_i, \cdot
\right] \right) \Psi \right\}  \; .
\end{eqnarray}
We note that such scalar masses have been considered in recent lattice simulations of 4 supercharge Yang-Mills in two dimensions to eliminate observed divergences in the scalars \cite{Kanamori:2008bk,Kanamori:2008yy,Kanamori:2009dk,Hanada:2009hq}.\footnote{The 1-loop considerations for this theory are similar to those of our maximally supersymmetric quantum mechanics, and therefore we expect the continuum two dimensional 4 supercharge theory should have a divergent partition function for thermal fermion boundary conditions. The scalar divergences in these lattice simulations are presumably a manifestation of this.}

The mass gives a quadratic potential for all the components of the matrix scalar fields $X_i$ which confines them to a region where $< X_i^2 > \, \sim \lambda / ( R \mu^2 )$ or smaller. Thus the classical moduli will also be confined to a region,
\begin{eqnarray}
R | \Delta x^{ab} | \sim \frac{1}{t^{1/2}} \frac{1}{R \mu}
\end{eqnarray}
and so the mass acts as a regulator with $\log r \sim \log (R \mu)$ in equation \eqref{eq:divaction}. Recall from our 1-loop arguments above that at low temperatures, for $R | \Delta x^{ab} | > \frac{1}{t^{1/2}}$ the diagonal non-constant modes and off-diagonal modes become weakly coupled, and it is this region that gives rise to the divergence in the partition function.
Thus we may regard the dimensionless $m \equiv (R \mu)$ as parameterizing the amount of this weakly coupled region allowed by the regulating mass.

Then our procedure is to compute the leading $N$ thermodynamic quantities for small fixed dimensionless  mass $m$ taking $N$ large enough to give the $N^2$ contribution of interest. Then the dimensionless mass  $m$ is reduced and the process repeated. The limit of zero mass should yield the correct finite $N^2$ contribution.

For this simple regulator unfortunately the subleading $N$ contribution apparently holds no physical interest. We note that there is another regulator that is more natural to use in this context, namely the mass term that preserves the full 16 supercharges of the theory where the quantum mechanics becomes the BMN plane wave matrix model \cite{BMN}. This mass term again lifts the non-compact quantum moduli space, rendering the thermal theory convergent. Now the theory \emph{with} regulator mass is dual to a known IIA closed string theory.
Indeed the supergravity vacuum solutions dual of this theory with mass term are known \cite{LLM}, although these are complicated by the fact that the mass term breaks the $SO(9)$ R-symmetry to $SO(3)\times SO(6)$, and correspondingly reduces the isometry of the dual supergravity solutions in the same manner. So far the thermal supergravity solutions have not been analysed and this would probably require a numerical treatment as one expects the metrics to be cohomogeneity two (for example using techniques such as \cite{Headrick:2009pv}). Thus in principle using this mass term one would obtain a well behaved thermodynamics and have a good string dual, even though the supergravity prediction is not yet known. Both the SYM and the gravity calculations are therefore interesting areas for further study. Unfortunately, just as the temperature must be small for the supergravity approximation to hold near the black hole horizon, it is likely that the regulator mass would also have to be taken to be small.

%
\section{Lattice implementation}
%

After integration over the fermions the continuum
Euclidean path integral, 
\beq
Z = \int dA dX \mathrm{Pf} \left( \mathcal{O} \right) e^{ - S_ 
{bos}}\eeq
is given by,
\begin{eqnarray}
S_{bos} & = & \frac{N}{\lambda} \mathrm{Tr} \oint^R d\tau
        \left\{ \frac{1}{2} (D_{\tau} X_i)^2 - \frac{1}{4} \left[ X_i,
X_j \right]^2 + \frac{1}{2} \mu^2 X_i^2 \right\}  \nonumber \\
\mathcal{O} & = & \gamma^\tau D_{\tau}  - \gamma^i \left[ X_i, \cdot
\right]
\end{eqnarray}
where $\mu$ parameterizes the scalar mass that we will use to regulate the thermal divergence.
The $\gamma^\tau, \gamma^i$ are a Euclidean representation of the Lorentzian Majorana-Weyl
gamma matrices, obeying $\{ \gamma^\mu , \gamma^\nu \} = \delta^{\mu\nu}$ for index $\nu = \{ \tau, i 
\}$. 
Thermal boundary conditions correspond to taking the
fermions antiperiodic on the Euclidean time circle and correspond
to a temperature $t=\lambda^{-1/3}/R$. 
The Pfaffian is in general complex \cite{Nicolai} giving rise to a potential `sign problem'. It is important
in principle to include the phase of the Pfaffian in the Monte-Carlo  
simulation, and we discuss this later.

We discretize this continuum model as,
\begin{eqnarray}\label{eq:actiondis}
       S_{bos} &=&  \frac{NL^3}{\lambda R^3}  \sum_{a=0}^{L-1} \mathrm
{Tr}\left[ \frac{1}{2}  \left( D_- X_i \right)_a - [ X_{i,a}, X_{j,a} ]
^2  + \frac{1}{2} m^2 X_{i,a}^2 \right] \nonumber \\
\mathcal{O}_{ab} & = & \left( \begin{array}{cc} 0 & (D_-)_{ab} \\ -(D_-)_
{ba} & 0 \end{array} \right) - \gamma^i \left[ X_{i,a}, \cdot
\right]  \mathrm{Id}_{ab}
\end{eqnarray}
where we have rescaled the continuum fields $X_{i,a}$ and $\Psi_{i,
\alpha}$
by powers of the lattice spacing $a=R/L$ where $L$ is the number of
lattice points to render them dimensionless. The dimensionless lattice mass regulator parameter $m$ is related to the continuum mass as $m = (R \mu)$, and as described above, it is this dimensionless $m$ that parameterizes how much of the weakly coupled divergent region is allowed by the regulator.

The covariant derivatives are given by
$(D_- W)_a = W_{i,a} - U_a W_{i,a-1} U_a^{\dagger}$
and we have introduced a Wilson gauge link field $U_a$. Notice that  
the fermionic operator is
free of doublers and is manifestly antisymmetric yielding a well defined Pfaffian on the lattice. 
An important subtlety is that in order to obtain this antisymmetric fermion operator after finite differencing we have chosen a twisted Euclidean representation for the Gamma matrices. In particular, we have taken a representation where,
$\gamma^{\tau} = i \left( \begin{array}{cc} 0 &+ \mathrm{Id}_8 \\ +
\mathrm{Id}_8 & 0 \end{array} \right)$. One may think of the theory as originating from a classical dimensional reduction of $\mathcal{N} = 1$ SYM in 10-dimensions. One can integrate out the fermions in 10-dimensional Lorentzian spacetime, continue to Euclidean time $\tau = i t$, and now take the theory to live on a 10-torus. In principle then one would dimensionally reduce this 10-torus to 1-dimension, $\tau$, by shrinking the spatial cycles to zero size. However instead our twisted theory is obtained by reducing to 1-dimension where the remaining dimension
corresponds to one of the space circles. If we had taken periodic boundary conditions for the fermions on all torus cycles one would expect no difference. However, we are imposing antiperiodic boundary conditions on one cycle. In principle this should be the time cycle. However, in our construction, when lifting back to 10-dimensions we have chosen it to be a space cycle. Thus the theory is physically different in our twisted case from the conventional untwisted theory.
However, since the difference between the theories is only seen in the fermion boundary conditions, we expect that both the twisted and untwisted theories share the same supergravity dual, which at leading order in the semiclassical approximation (ie. large $N$ in the field theory) has no contribution from fermions, beyond the restriction on the possible manifolds from spin structure. Presumably differences in the twisted and untwisted theories therefore show up at subleading order in $N$, something that we do not study here, and from the closed string perspective must arise when one considers the backreaction of the supergravity fermions.

In a super-renormalizable theory we expect only a finite number of divergences can arise in our lattice theory. Potentially such divergences may 
induce a RG flow away from the maximally supersymmetric continuum limit of interest. However, naive expectation is that for quantum mechanics, there will be no divergences. When fermions are present this is too quick a conclusion, as fermion loops arising from a 2-fermion 2-boson interaction or fermion tadpole loops suffer a superficial divergence on the lattice if kinetic terms do not respect parity (as in the continuum) \cite{Giedt}. However, as argued in \cite{four}, in the Yang-Mills case
there is no 2-fermion 2-boson interaction, and the possible fermion tadpole loop does exist, but vanishes due to the gauge structure. Thus we expect the lattice action is finite and hence will flow {\it without fine
tuning} to the correct maximally supersymmetric continuum theory with decreasing lattice spacing. In \cite{Catterall:2008yz} this was tested by simulating the theory with periodic boundary conditions and testing whether the path integral, in this case the Witten index, was dependent on $R$ in the continuum limit.

We use the RHMC algorithm \cite{rhmc} to sample configurations using
the absolute value of the Pfaffian. The phase may be
re-incorporated in the expectation value of an observable ${\cal A}$
by reweighting as $<\mathcal{A}> = \frac{\sum_{m} \left( \mathcal{A}  
e^{i \phi}
\right)}{\sum_{m} \left( e^{i \phi} \right)}$.
Here $e^{i \phi(\mathcal{O})}$ is the phase of the Pfaffian
and the sum runs over all members of our phase quenched ensemble.

%
\section{Lattice Results}
%

In this section we shall report various properties of the thermal supersymmetric quantum mechanics discussed above which we have simulated on the lattice over the temperature range  $0.3 < t < 5$.
We will report on various observables. 
The mean energy, $\epsilon$ and the absolute value of the trace of the 
Polyakov loop, $P$, are given in terms of our lattice action above as, 
\begin{eqnarray}
<\epsilon /t> & = & \frac{3}{N^2}\left(\frac{9}{2} L (N^2-1)-
<S_{bos}>\right) \nonumber \\ 
P & = & \frac{1}{N}<|\mathrm{Tr}\prod_{a=0}^{L-1}U_a|> 
\end{eqnarray}
as discussed in \cite{four}.
We also compute the Pfaffian phase $\phi$, and the width of the scalar eigenvalue distributions, $W$, given in terms of the eigenvalue distribution $p(x)$ of the scalars $X_i$, as,
\begin{eqnarray}
W & = &  \beta^{-3/4} \int dx \; |x| \; p(x)
\end{eqnarray}
The inclusion of $1/N^2, 1/N$ in the above definitions is
to ensure these quantities are finite in the t'Hooft limit for a deconfined phase. 

We begin by examining how many lattice points will be required by our lattice approximation to recover reasonable continuum predictions for the observables we are interested in. We will assume that continuum systematic errors are similar to those in the quenched theory. 
Since the quenched theory is easy to simulate we may assess these errors. In figure \ref{fig:quenched} we show a plot of energy over temperature against temperature for various numbers of lattice points. We see that already for 5 lattice points the systematic error appears to be small compared to the statistical uncertainty in the Monte Carlo method. This is true for all observables discussed in this section, and over the whole temperature range studied. 

Hence in the remainder of this section we will largely present results for 5 point lattices. Whilst in the quenched theory we have the luxury of being able to calculate for larger lattices, in the unquenched case going beyond 10 lattice points represents a considerable challenge. Hence it is very encouraging that already interesting information is obtained from 5 points. However, one must bear in mind that other observables that probe more local properties of the theory may be more sensitive to continuum systematic error. We leave the very interesting question of going beyond the crude thermodynamic observables we are examining here to future work.

Another use for the quenched theory is to assess how quickly the `t Hooft scaling sets in. In the same figure we plot the energy over temperature for $N = 5, 12, 30$. As for pure Yang-Mills, we see that the large $N$ scaling sets in quickly, with $SU(5)$ already providing quite a good approximation to the infinite $N$ extrapolation. For the full theory we will be able to present data for $N$ up to $12$, and for our variables of interest, this is reasonable. However, as we shall see later, the issue of the thermal divergence means that going to larger $N$ may be necessary in future work.

\FIGURE[h]{
\centerline{
\includegraphics[height=2.7in,width=5.0in]{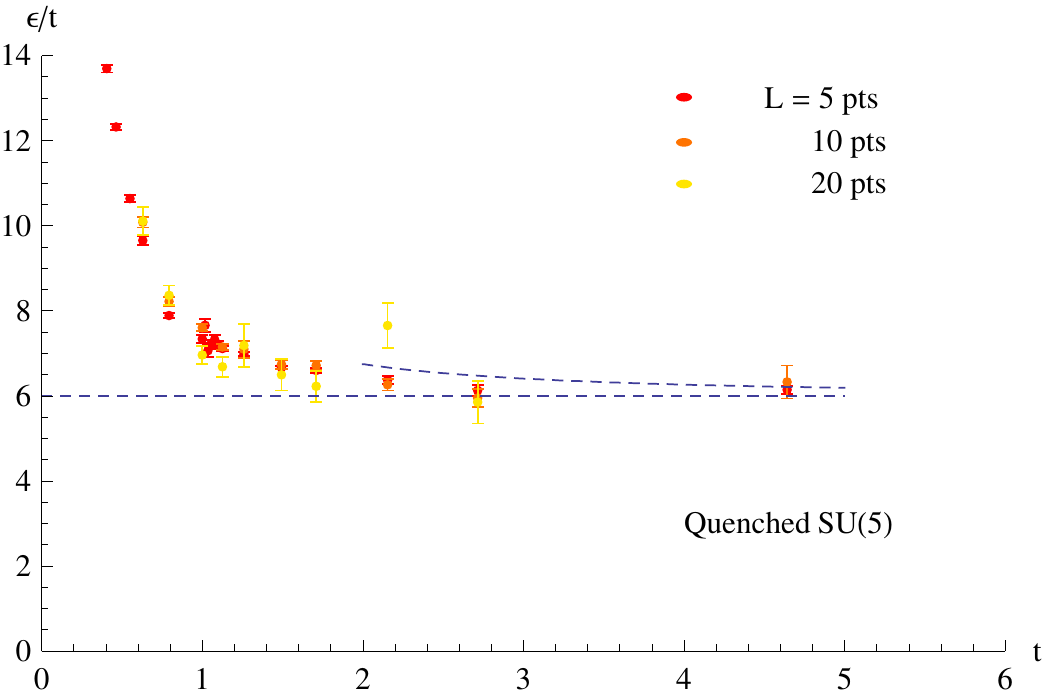}
}
\centerline{
\includegraphics[height=2.7in,width=5.0in]{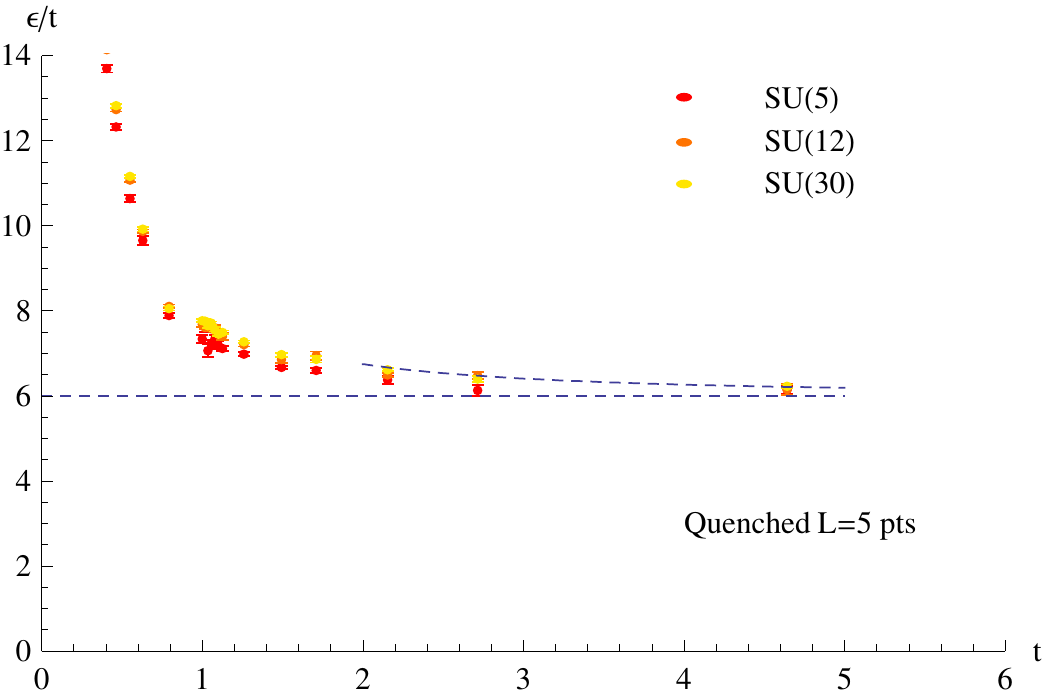}
}
\caption{
Plots of $\epsilon/t$ against $t$ for the quenched theory. The top plot shows $SU(5)$ for various numbers of lattice points, indicating that the continuum is quickly reached over the temperature range considered. The bottom plot shows the approach to the 't Hooft limit, for $N = 5,12, 30$ all for 5 lattice points.
}
\label{fig:quenched}
}

We have discussed the claimed IR thermal divergence of the quantum mechanics due to its exact Lorentzian moduli space. A natural question is how this manifests itself on the lattice. If the partition function fails to exist, then presumably the Monte Carlo scheme must also fail in some manner.
Choosing an appropriate gauge, the divergence is associated with the classical flat directions of the potential for the diagonal components of the scalar matrix fields. The integration over these diagonal modes should give a divergence when the diagonal values become large and well separated. A gauge invariant observable that is sensitive to this are the eigenvalues of these scalar matrices. We observe in certain simulations that the maximum value of the scalar eigenvalues starts to increase in a seemingly unbounded manner under Monte Carlo iteration time. This appears to be in accord with the Monte Carlo method beginning to sample the range of field space where the divergence originates from. Certainly at this point the Monte Carlo procedure can no longer practically provide statistically independent configurations, and hence the procedure breaks down.  

For low temperatures we always observe that the eigenvalues increase unboundedly almost immediately the simulation is started. For temperatures $t \simeq 1.0$ we observe that the simulations must be run for many configuration times in order to see the divergence start. In figure \ref{fig:diverge} we illustrate this by showing two typical maximum eigenvalue trajectories for 5 point $N = 5$ Monte Carlo runs with $t = 0.94$ and $t = 1.12$. As is typical for $t < 1.0$ the eigenvalues diverge very quickly. For the instance of the larger temperature shown we must run for 600 configurations in order to see the divergence. Presumably the Monte Carlo initially scans the strongly coupled region and tunnels out of this to the divergent region, and this becomes less probable in a given update step as the temperature is increased.
For larger $t > 1.3$ for $N = 5$ we haven't seen a divergence in any of our runs starting with initial configurations where the eigenvalues are tightly clustered, although presumably running for very large numbers of configurations one would. Thus the observed lattice eigenvalue run away which invalidates
 the Monte Carlo scheme appears to be consistent with our formal earlier discussion of the thermal divergence. We note that, as one would expect, in the quenched theory we see no such eigenvalue run away behaviour in the lattice simulation. For the periodic theory we also saw no eigenvalue divergence \cite{Catterall:2008yz}.

\FIGURE[h]{
\centerline{
\includegraphics[height=2.7in,width=5.0in]{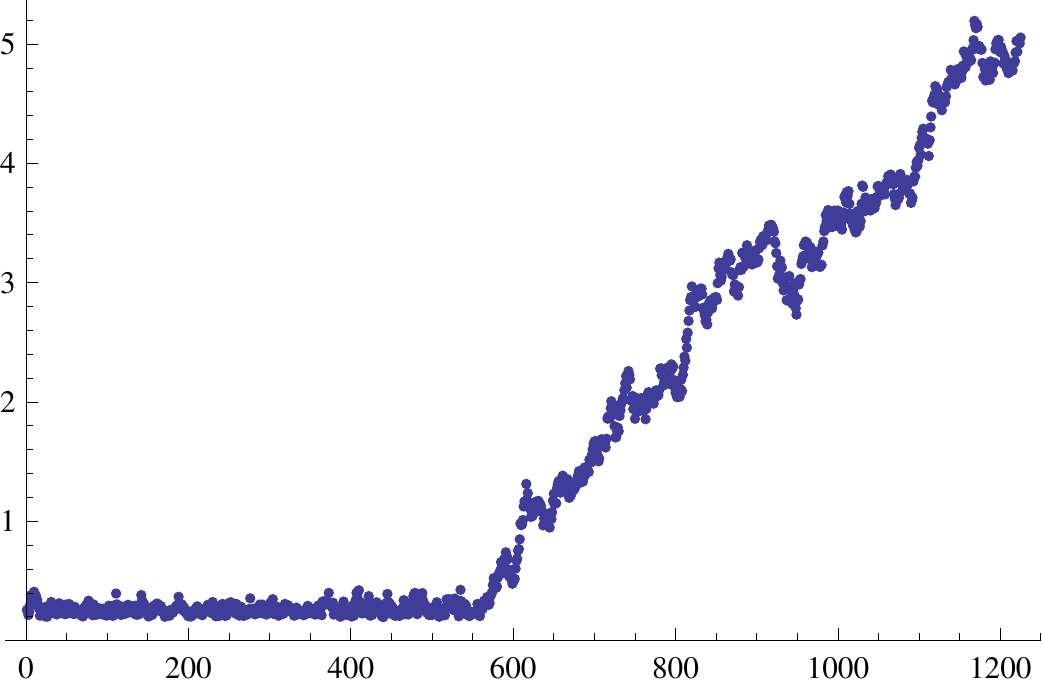}
}
\centerline{
\includegraphics[height=2.7in,width=5.0in]{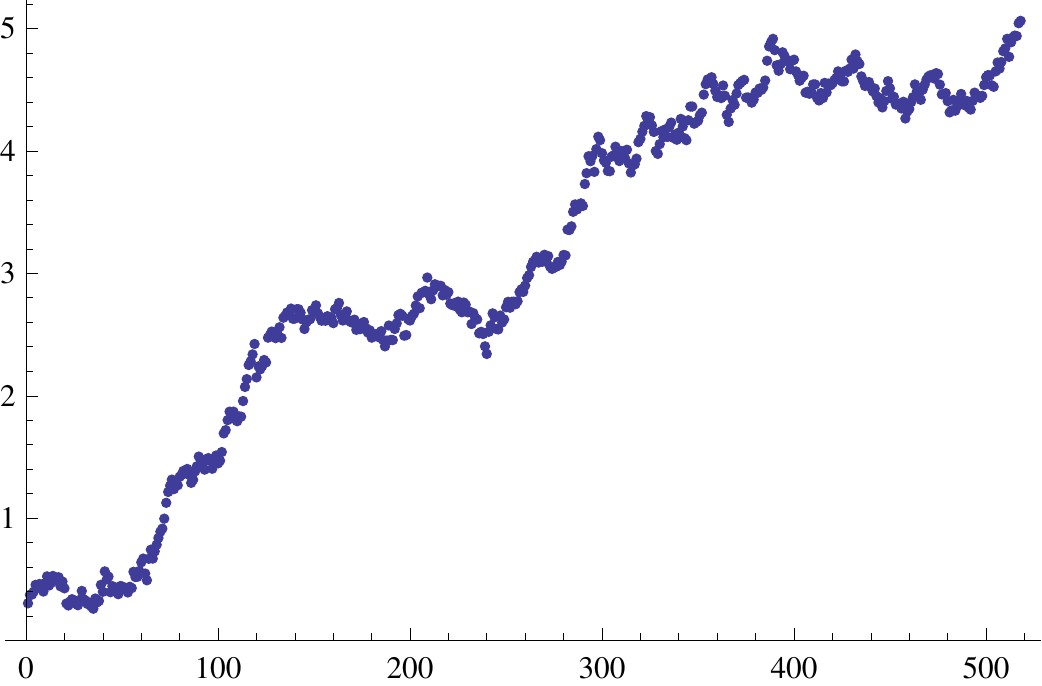}
}
\caption{
Plot of the maximum absolute value of the scalar eigenvalues against Monte Carlo configuration number, for $N = 5$ with 5 lattice points with the top plot having $t = 1.12$ and the bottom plot having $t = 0.94$. These represent typical Monte Carlo sequences and show that for lower temperatures the divergence quickly sets in, while for higher temperatures it may take many configurations before the instability is seen, with $t \simeq 1.0$ marking the divide in behaviour.
}
\label{fig:diverge}
}

It is interesting that for high temperatures, one may use the Monte Carlo to sample the strongly coupled region of the partition function for many configurations before one encounters the divergent region and the Monte Carlo breaks down. This is essentially what allowed the previous studies \cite{Catterall:2008yz, Nishimura16} to display data for the thermal model without explicitly regulating the divergence. In figure \ref{fig:meta} we use this `metastability' to plot the energy over temperature for the unregulated theory in the phase quenched approximation. The solid small data points are ones where no divergence was encountered during the runs. The open squares are points where divergences were encountered after a number of configurations, and the data for the Monte Carlo series is truncated by hand to the region before the divergence. We see that as $N$ is increased we find the temperature at which the instability appears to set in decreases, although not substantially. Note however that as $N$ increases the number of configurations required to obtain the same level of statistical error in a measurement also decreases, and there one typically is using fewer configurations, so perhaps this is not surprising. We also plot data for 10 point $N = 3,5$, where again we may obtain data for lower temperatures mainly by virtue of the fact that fewer configurations are required to obtain a given level of statistical error.

We display these curves for comparison with the previous work \cite{Catterall:2008yz, Nishimura16} and not because this ad hoc method of truncating Monte Carlo series by hand should be taken seriously as an algorithm. Still, it is very interesting that the energy curves obtained appear to be consistent with the low temperature supergravity prediction, also shown in the figures. Thus it appears naively that the strongly coupled region of the field space where the Monte Carlo sampling works is indeed responsible for the black hole behaviour, and the field space associated with the divergence, whilst physically important, appears to have little impact on this black hole behaviour. It would be very interesting if this could be quantified more precisely

\FIGURE[h]{
\centerline{
\includegraphics[height=2.7in,width=5.0in]{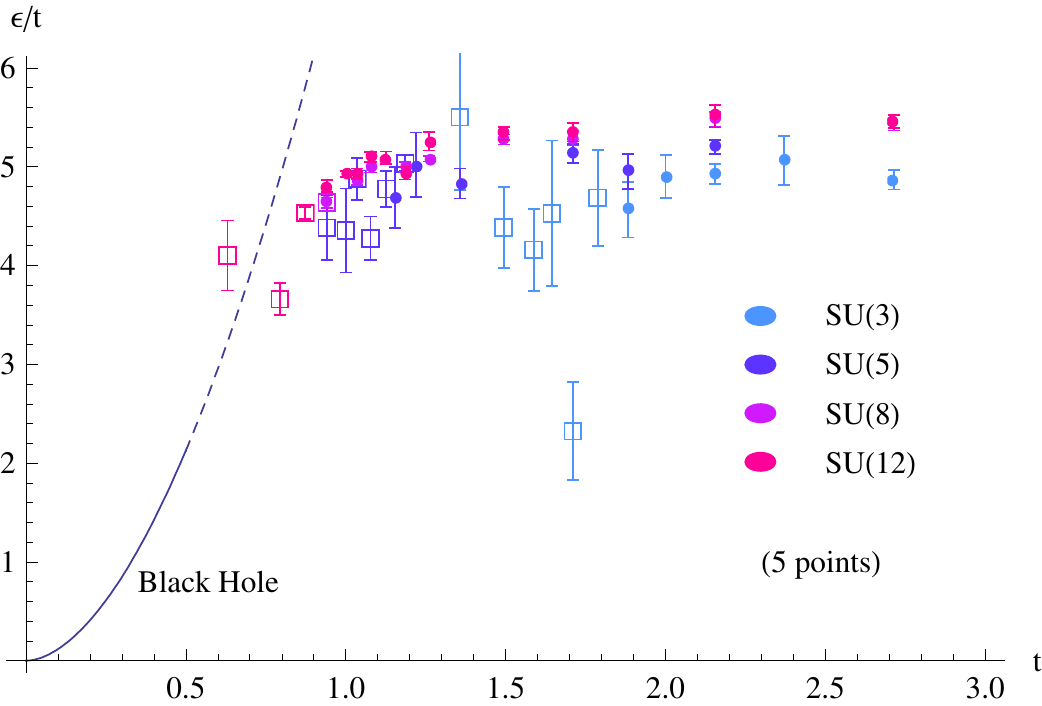}
}
\centerline{
\includegraphics[height=2.7in,width=5.0in]{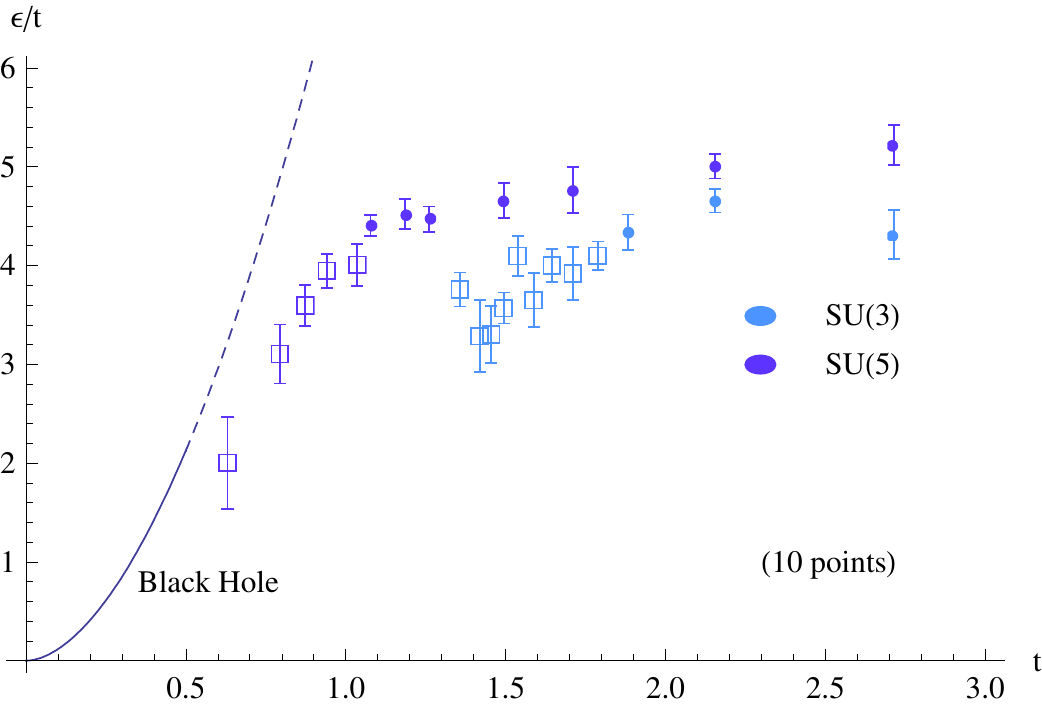}
}
\caption{
Plots of $\epsilon/t$ against $t$ for various $N$ computed without a regulator mass for 5 and 10 lattice points, but instead artificially truncating the Monte Carlo sequences when the scalar eigenvalues start to diverge. The open squares denote points where truncation was required, small discs are points where no divergence was seen for the number of configurations used. We note that for larger $N$ or more lattice points the error bars are smaller for fewer configurations and hence one may probe to lower temperatures. Whilst the method is clearly ad hoc and difficult to justify, the results do look plausibly in agreement with the predicted dual black hole low temperature behaviour, plotted as the solid blue curve. Configurations are generated in the phase quenched approximation. 
}
\label{fig:meta}
}

Having now explored the behaviour of the divergence with Monte Carlo
simulations and in particular
what happens if we crudely ignore it, in the remainder of the section we shall treat it properly and regulate it with a scalar mass term. In figure \ref{fig:evalsmass} we plot all the eigenvalues of the scalars against Monte Carlo RHMC time in the mass regulated theory, for masses $m = 0.05, 0.1, 0.5$ for a 5 point lattice for $N = 12$ and a quite small temperature, $t = 0.43$. We see that rather than diverging, the eigenvalues now appear bounded, with a core dense region of approximately the same width for the various $m$, surrounded by a more diffuse `halo' of eigenvalues who's extent increases as $m$ decreases ie. as the regulator is removed.

We note that for the low $m$ runs, the motion of the `halo'  eigenvalues is very slow in Monte Carlo time, and hence their dynamics is not correctly treated by our algorithm, where a configuration is taken for sampling at unit intervals of the RHMC time. However, since it is the strongly coupled core eigenvalues that give rise to the behaviour of interest, we will not discuss this concern further here, but merely note that it is clearly a technical issue for future study to better treat the split of light and heavy degrees of freedom in this system.

\FIGURE[h]{
\centerline{
\includegraphics[height=2.1in,width=4.0in]{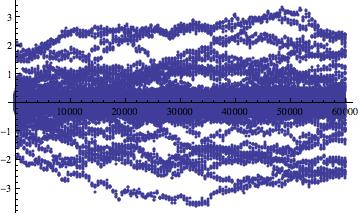}
}
\centerline{
\includegraphics[height=2.1in,width=4.0in]{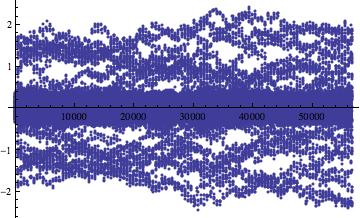}
}
\centerline{
\includegraphics[height=2.1in,width=4.0in]{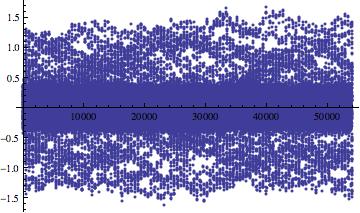}
}
\caption{Plot of scalar eigenvalues for $SU(12)$ at $t = 0.43$ with 5 lattice points as function of Monte Carlo configuration time for regulator masses  $m = 0.05$ (top), $0.1$ (middle) and $0.2$ (bottom). Congurations are generated in the phase quenched approximation.}
\label{fig:evalsmass}
}

With a regulated partition function a key question to address is whether the fluctuations of the Pfaffian phase present an obstacle to a numerical approach.  Using the resources available currently we have only computed this phase for 5 point lattices with $N \le 5$ as the Pfaffian calculation is very costly computationally. However the results are very encouraging. In figure \ref{fig:Pfaff} we plot the expectation value for the cosine of the Pfaffian phase against temperature for $N = 3$ and $5$ for 5 point lattices for 3 values of the regulator $m$. 
It is not surprising that at high temperature the phase vanishes, and the thermal mass given to the fermions presumably suppresses the dynamics of the phase. However, initially more surprising is that at low temperatures we observe that the phase also becomes trivial. This is far less expected. 

Also interesting is that as the regulator mass is decreased, the phase appears to become closer to zero. We have no reason to expect the value to be trivial in the continuum limit, but apparently it is rather small over the temperature range probed. One point to note is that the phase appears closer to zero for $N = 3$ than $N = 5$. One might worry that increasing $N$ leads to greater phase fluctuations.  Since we have not simulated the phase for greater values of $N$ we cannot comment concretely on this, suffice to say that since the phase is already small for $N = 5$, provided it does not increase by much for larger $N$, it still would not be problematic.

In the remainder of this section we will present results calculated in the phase quenched approximation, as for $N = 8, 12$ we haven't Pfaffian phase data. However we note that for $N = 3, 5$, we may re-weight observables by the phase and when we do this for the observables shown, as one would expect from the results in figure \ref{fig:evalsmass}, the change in the value of the observables due to re-weighting is insignificant compared to the statistical error in the results, and cannot be seen easily by eye.

\FIGURE[h]{
\centerline{
\includegraphics[height=2.0in,width=4.0in]{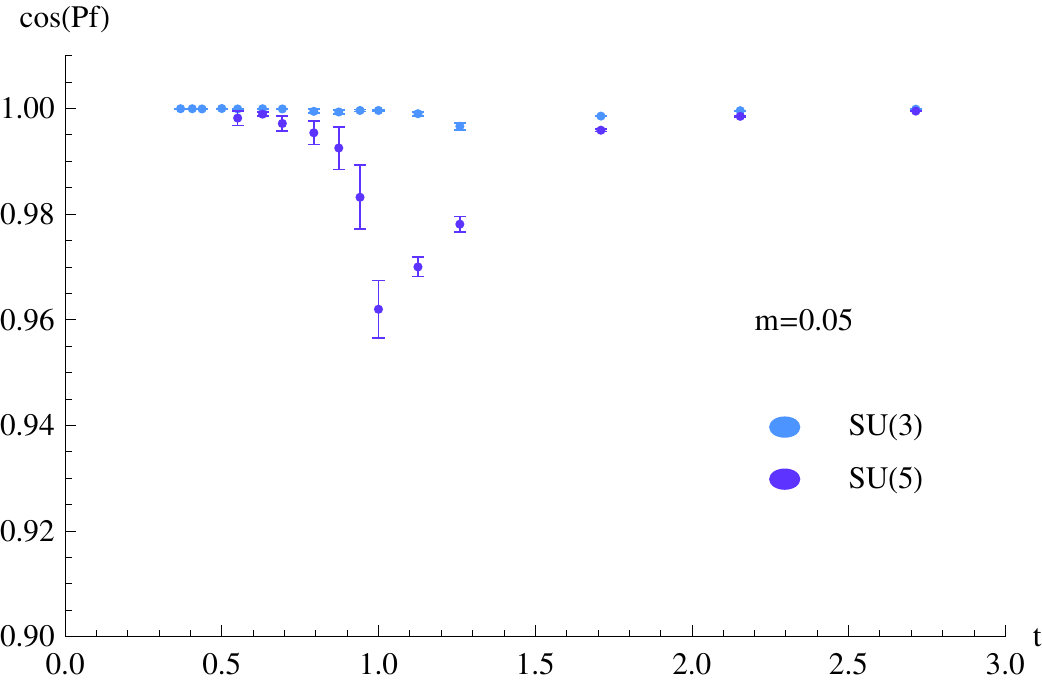}
}
\centerline{
\includegraphics[height=2.0in,width=4.0in]{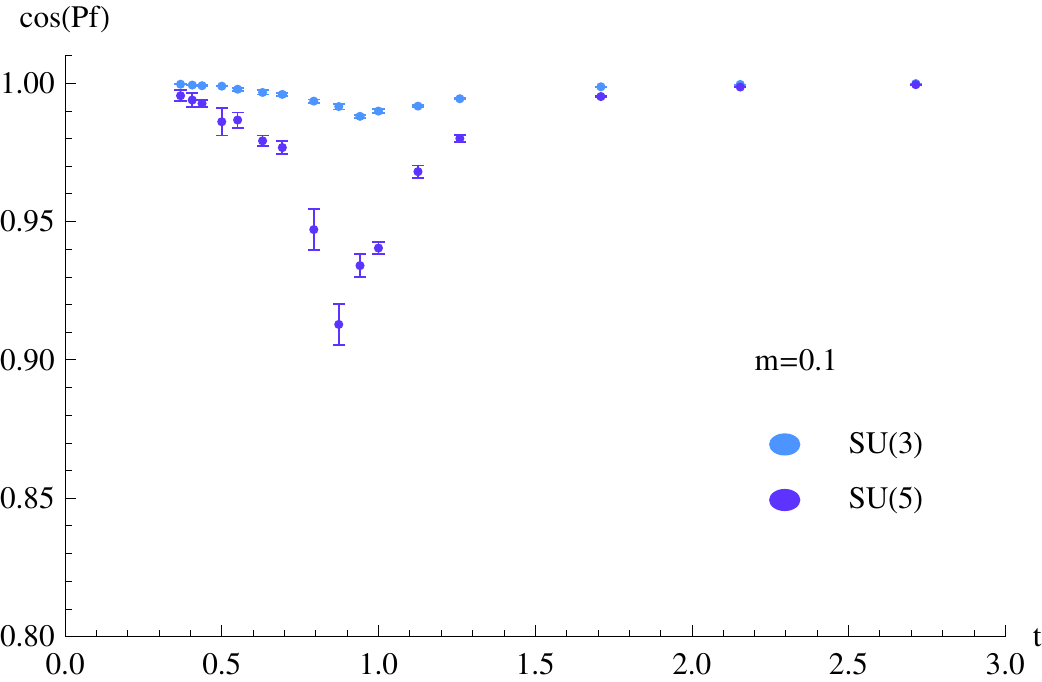}
}
\centerline{
\includegraphics[height=2.0in,width=4.0in]{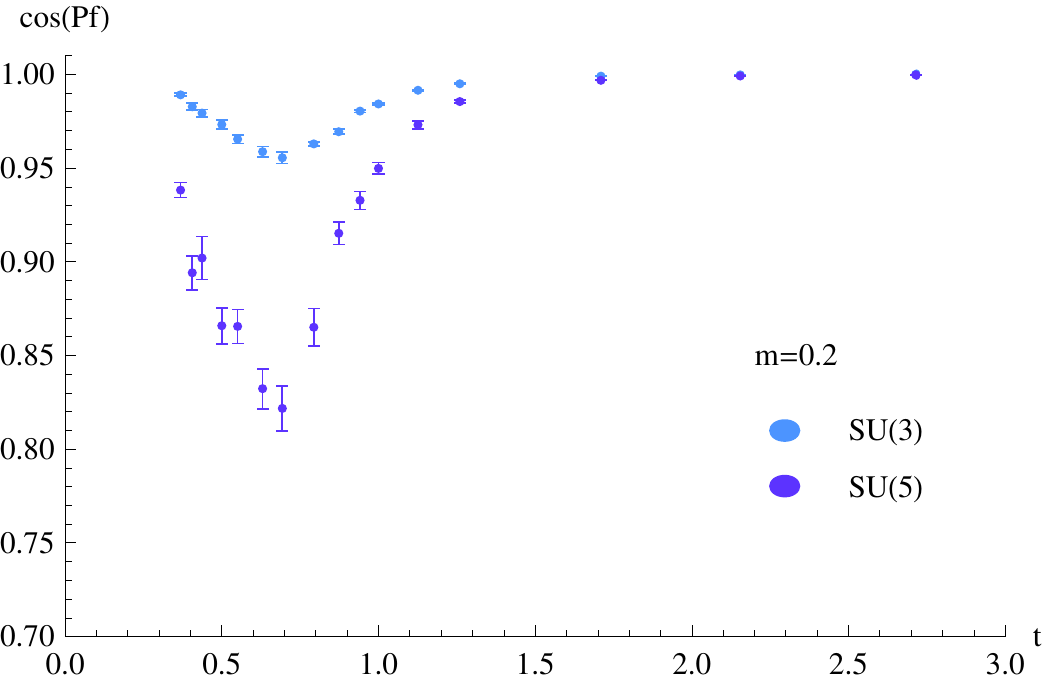}
}
\caption{
Plot of the average cosine of the Pfaffian phase for $N = 3$ and $5$ for 5 lattice points for regulator masses  $m = 0.05$ (top), $0.1$ (middle) and $0.2$ (bottom). We note that at low and high temperatures the Pfaffian phase appears to play little dynamical role. Its effect appears to increase as the regulator is removed, and as $N$ is increased, although we always observe it to be rather small. Reweighting other observables shown in later plots for $N = 3, 5$ gives no discernable difference from the phase quenched approximation.
}
\label{fig:Pfaff}
}

Having regulated the thermal theory, examined the continuum limit and checked the effect of the Pfaffian phase, we are ready to show data for observables of physical interest. In figure \ref{fig:ergmass} we show the energy over temperature plotted against dimensionless temperature for 3 values of the regulator. In figure \ref{fig:polymass} we show the trace of the Polyakov loop and scalar eigenvalue width. Let us now focus on the energy curves. We expect to see a dependence on the regulator $m$, and indeed this is evident. For larger $m$ the large $N$ scaling sets in earlier, as the leading $N^2$ contribution to the energy is picked out from the subleading regulated divergence more easily. Hence the curves for $N = 5, 8, 12$ are all rather close. However, of course it is the large $N$ limit curve for small $m$ that we are actually interested in. We see that for our smallest $m$, as expected, one must go to higher $N$ to see the large $N$ scaling. Hence whilst $N = 8, 12$ are close, $N = 5$ is now quite different. Hence as the regulator is further removed, presumably increasingly large $N$ is required to extract the leading $N^2$ behaviour.

The curves for $m = 0.05$ give our best data for the theory as the regulator is removed. We see that the curves appear to interpolate between zero energy at zero temperature and $\epsilon$ being proportional to $t$ at large $t$ as we expect for quantum mechanics.\footnote{We note that for the larger $m$ curves we do not see $\epsilon \sim t$ as we are actually scaling the regulator mass with temperature; $\mu = m R$.} We also see that as $N$ is increased, and as the regulator is removed, the curves appear to approach the supergravity low temperature prediction, but the fit is visually still not that good for $m = 0.05$. This should be contrasted with the same data for the quenched theory shown in figure \ref{fig:quenched}, where the energy behaviour is totally different.

\FIGURE[h]{
\centerline{
\includegraphics[height=2.in,width=4.0in]{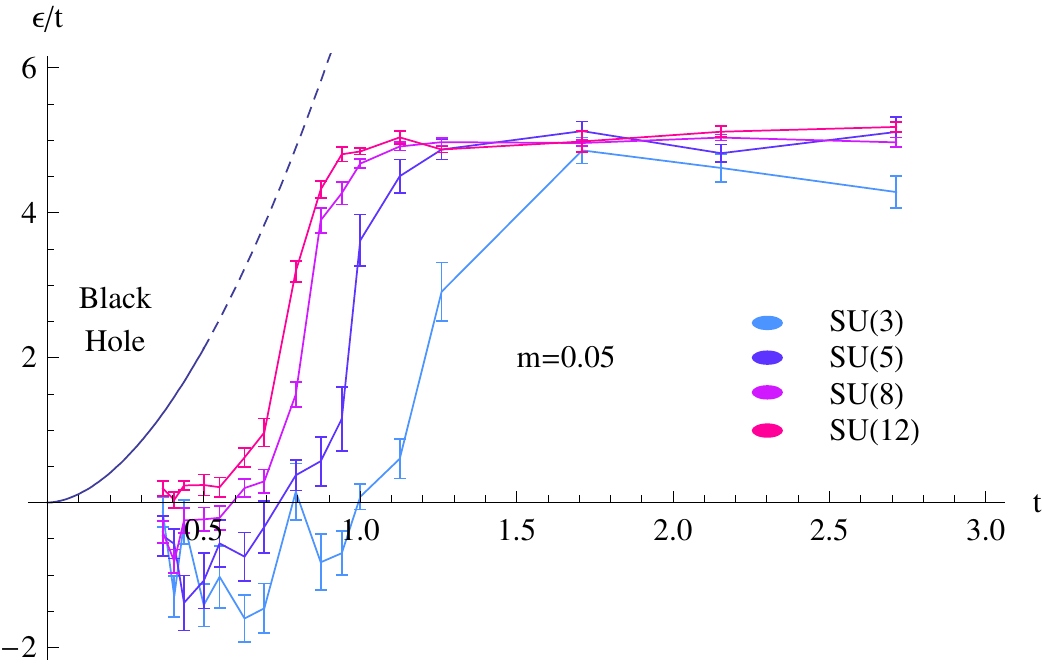} 
}
\centerline{
\includegraphics[height=2.in,width=4.0in]{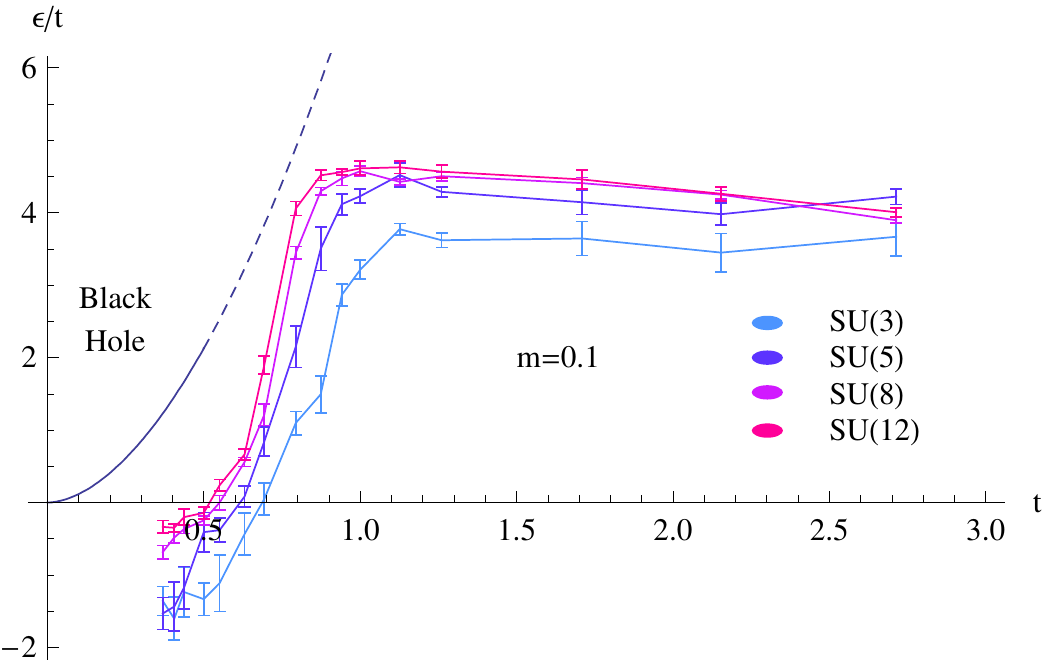} 
}
\centerline{
\includegraphics[height=2.in,width=4.0in]{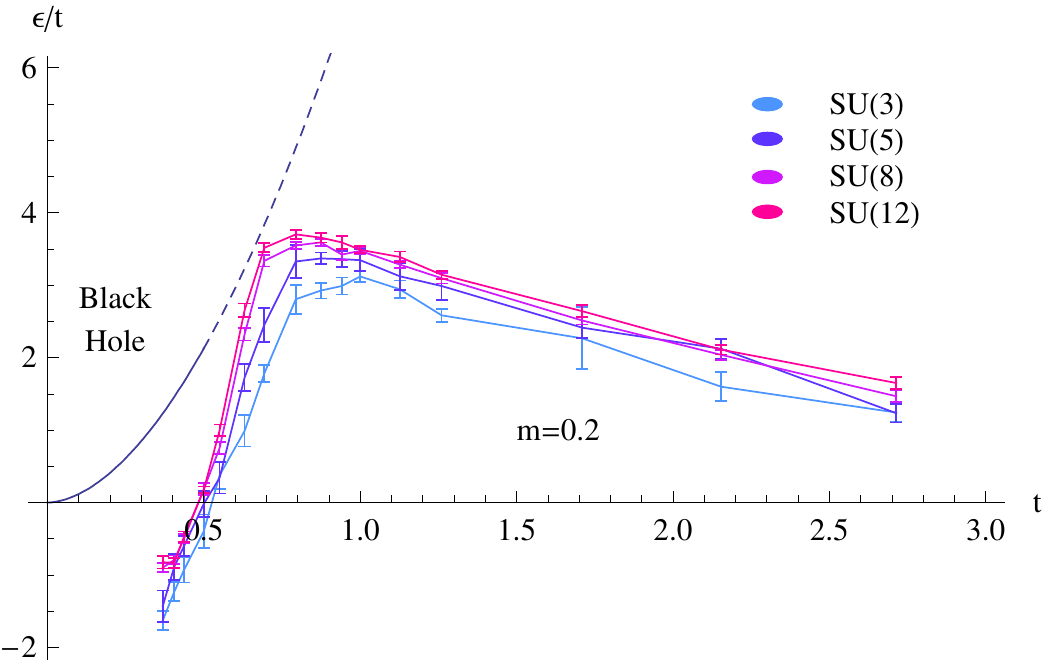} 
}
\vspace{0.0in}
\caption{Plots of  $\epsilon/t$ verses $t$ for $N = 3, 5, 8$ and $12$, with the regulator mass being $m = 0.05$ (top),  $ 0.1$(middle) and $0.2$ (bottom). We observe that for the larger regulator mass the various $N$ produce similar results, indicating that small $N$ are already close to the large $N$ limiting behaviour. For smaller regulator mass we observe that larger $N$ are required to find the limiting large $N$ curve, as expected, as the divergent $O(N)$ part of the action is relatively larger the lower the regulator mass.  These results are calculated in the phase quenched approximation for $N = 8,12$ and for $N = 3,5$ phase reweighting makes no discernable difference.
}
\label{fig:ergmass}
}

\FIGURE[h]{
\centerline{
\includegraphics[height=2in,width=3.0in]{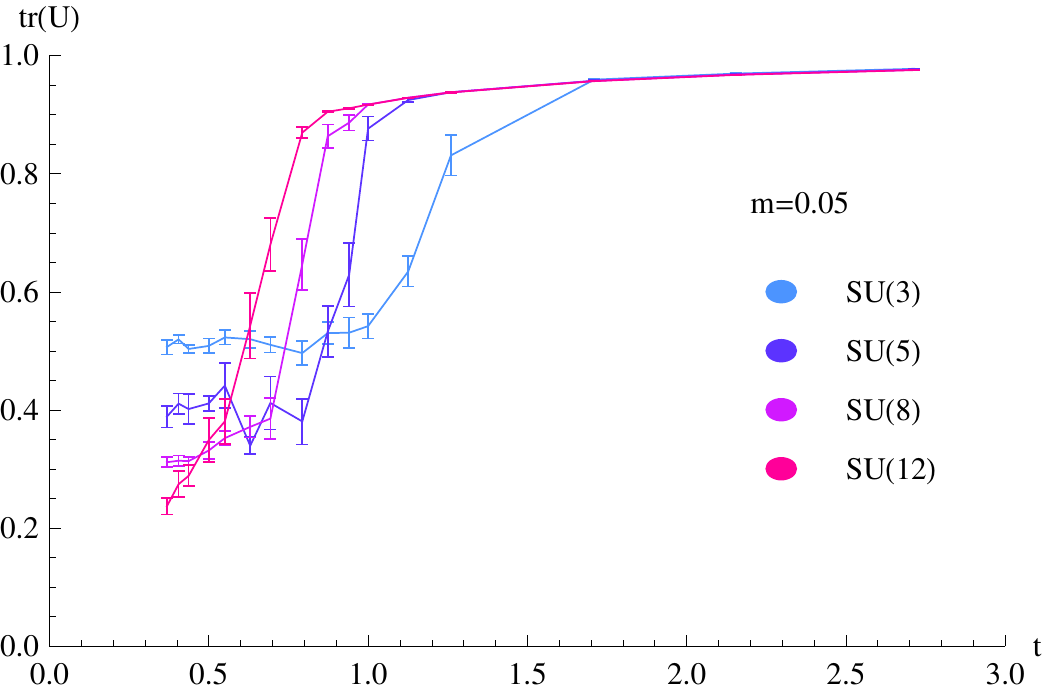} \includegraphics[height=2in,width=3.0in]{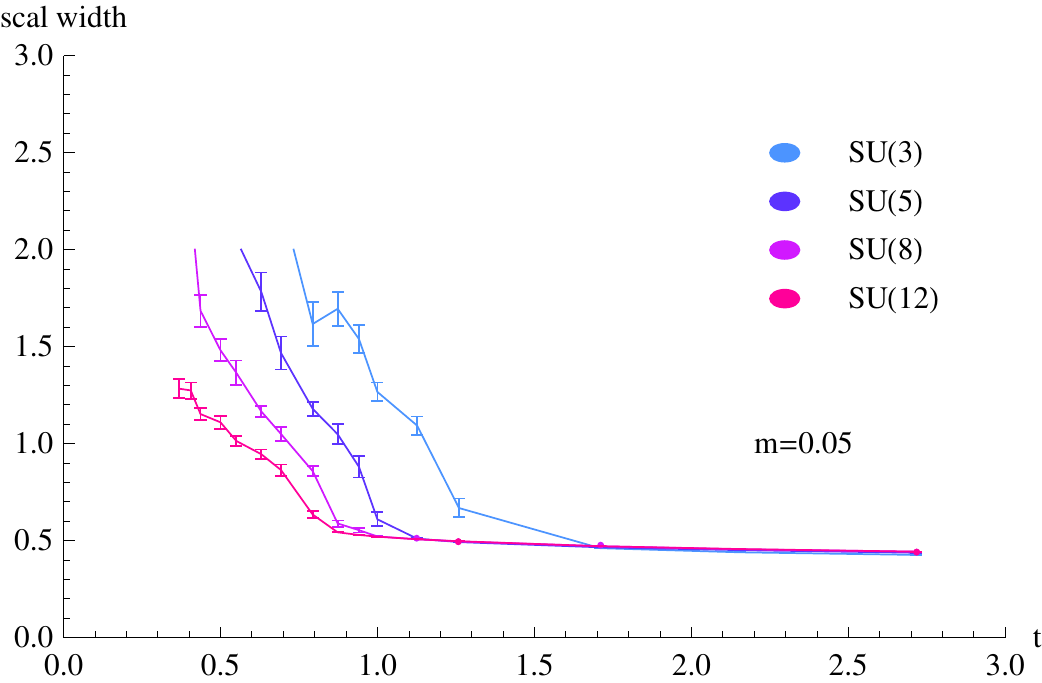}
}
\centerline{
\includegraphics[height=2in,width=3.0in]{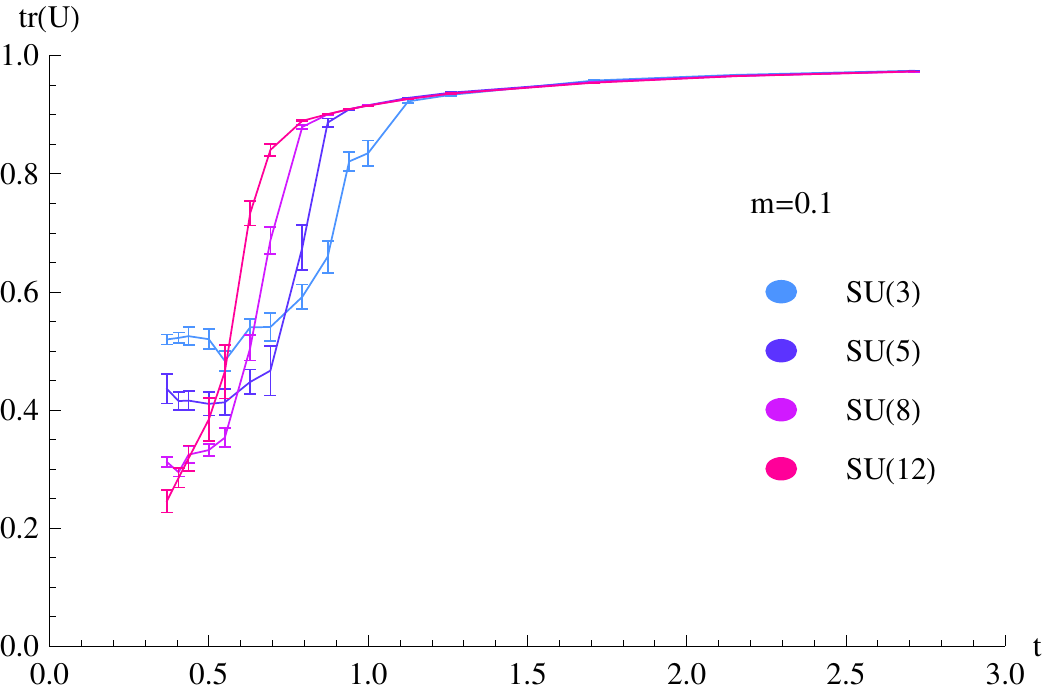} \includegraphics[height=2in,width=3.0in]{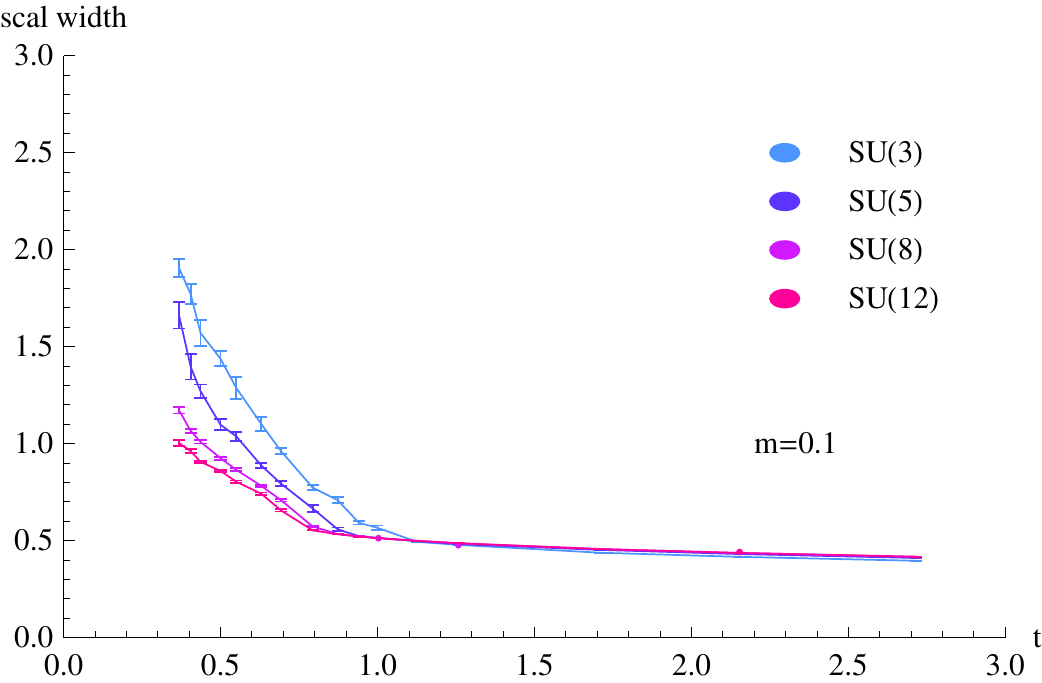}
}
\centerline{
\includegraphics[height=2in,width=3.0in]{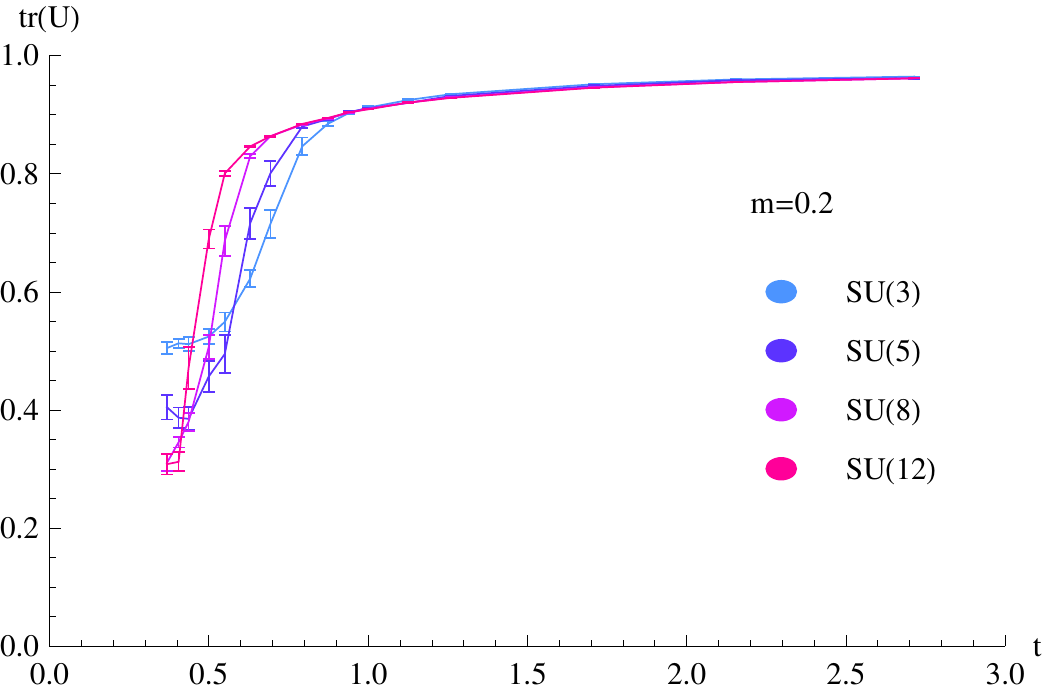} \includegraphics[height=2in,width=3.0in]{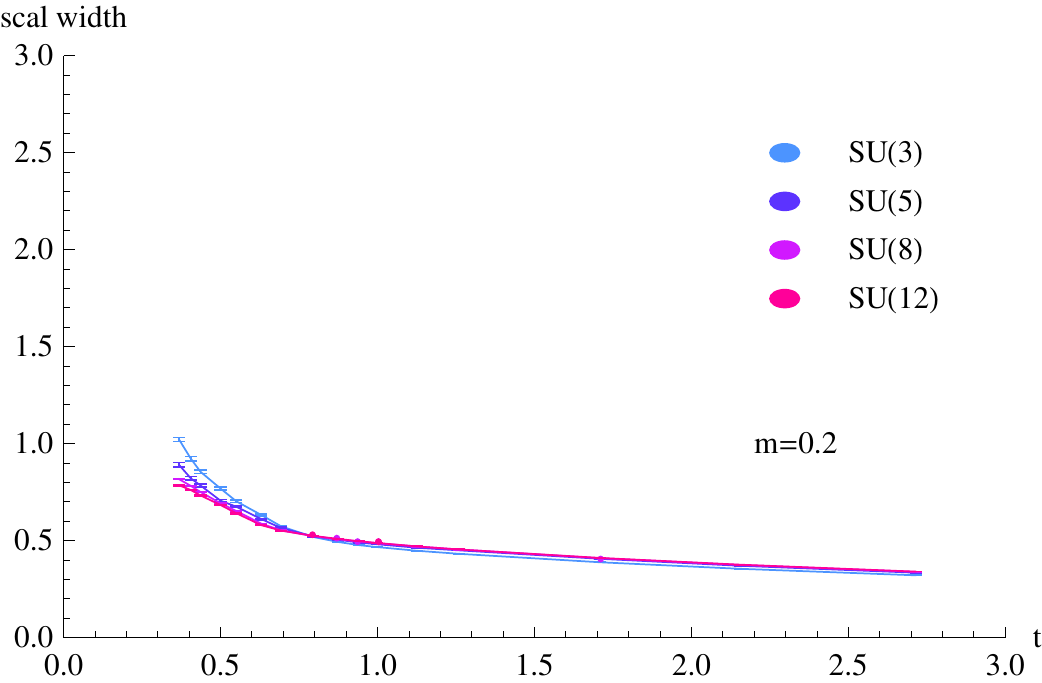}
}
\vspace{0.0in}
\caption{
Left plots show the Polyakov loop verses $t$ for $N = 3, 5, 8$ and $12$, with the regulator mass being $m = 0.05$ (top),  $ 0.1$(middle) and $0.2$ (bottom). Right plots show the scalar eigenvalue width.
}
\label{fig:polymass}
}

%
\section{Summary}
%

In this paper we have taken a more careful look at numerically simulating maximally supersymmetric gauged quantum mechanics, the worldvolume theory of $N$ D0-branes at finite temperature, which is conjectured to be dual to a closed string theory that includes quantum gravity. In particular for large $N$ and low temperatures, the thermodynamic behaviour of this quantum mechanics is supposed to reproduce the microscopic entropy of certain supergravity black holes. Since this theory is strongly coupled in the IR, it seems likely that a numerical approach is the only way to solve it.

We have emphasized that the exact quantum moduli space of the Lorentzian theory \cite{Sethi} or corresponding continuum of states down to zero energy \cite{deWit} appears to render the thermal partition function divergent. To our knowledge, such a conclusion was first claimed to give a thermal divergence in \cite{Kabat2}. 
Following earlier work \cite{Nishimura16} we have used the 1-loop approximation to lend further weight to the claim, arguing that the thermal partition function is divergent for any temperature and $N$, and being careful to consider the nature of the leading quantum corrections to the classical moduli space from all fluctuations. 
Whilst a thermal potential is generated on the bosonic moduli space, the potential is not strong enough to render the partition function convergent. We believe this divergence is dual to the Hawking evaporation of the D0-branes from the black hole, and the effect is subleading in $N$ to the expected $N^2$ finite behaviour of the free energy. The instability can be hard to see in Monte Carlo simulation.
Previous simulations of this thermal quantum mechanics saw signals of this instability but assigned it either to a lattice artefact \cite{Catterall:2008yz} or to a divergence arising only at temperatures below $T \sim 1/N$ \cite{Nishimura16}. Nonetheless such an instability renders such Monte Carlo simulations formally ill-defined.

In order to proceed with simulation we add a regulator mass to the scalars, and simulate the theory. We have argued that since the divergence is subleading in $N$, we may consider taking the large $N$ limit to extract the finite behaviour of the free energy which leads in $N$, and then remove the regulator. We have performed preliminary simulations where this is done, and the procedure appears to function as expected. We obtain results that appear roughly consistent with a low temperature behaviour predicted by a semiclassical black hole analysis when the regulator is removed, 
but the quality of the data should be considerably improved taking larger $N$ and smaller regulator mass in future work to draw any firm conclusions that test this holographic correspondence.

The `sign problem' that arises from the Pfaffian phase in the thermal theory is potentially dangerous to Monte Carlo simulation. We have carefully computed it for $N = 3, 5$, and find that it is reassuringly small over the interesting range of temperatures simulated. If this had not been the case, it would have spelt disaster for future simulations of thermal 16 supercharge Yang-Mills. Thus it is extremely encouraging that this possible difficulty does not appear to present itself in practice.

%
\section*{Acknowledgements}
%

We have benefitted greatly from numerous discussions with Matthew Headrick. We also thank Oleg Lunin, Kostas Skenderis and Mithat Unsal for useful discussion.
SC is supported in
part by DOE grant
DE-FG02-85ER40237. TW is supported by a STFC advanced
fellowship and a Halliday award. Simulations were performed using
USQCD resources at Fermilab.

\appendix

\section{Thermal 1-loop calculation}
\label{app:KK}

We will consider the quenched model first, and then discuss the theory with the fermions.
We may decompose the fields into diagonal components and off-diagonal ones.
\begin{eqnarray}
A(\tau) & = & a(\tau) + \hat{A}(\tau)  \nonumber \\
X_{i}(\tau) & = & x_i(\tau)  + \hat{X}_{i}(\tau) 
\end{eqnarray}
so that $a, x_i$ are diagonal and $\hat A, \hat X_i$ are off-diagonal. Now we have included the diagonal Kaluza-Klein modes as the non-constant parts of $a(\tau)$ and $x_i(\tau)$. As discussed in the main text the classical moduli are the constant diagonal modes, so that a classical vacuum configuration is defined by $\hat{A} = 0, \hat{X}_i = 0$ and $a(\tau), x_i(\tau)$ being constant in time.
Then we may write the action as,
\begin{eqnarray}
S & = & \frac{N}{\lambda}  \oint^R d\tau \,
      \tr  \left(  \frac{1}{2} \dot{x}_i^2 + \frac{1}{2} \hat{X}_\mu  \left( \hat{D}_\mu \hat{D}_\nu - 2 [ \hat{D}_\mu, \hat{D}_\nu ] - \delta_{\mu\nu} (\hat{D}_\alpha)^2 \right) \hat{X}_\nu + O(  \hat{X}^3 ) \right)
      \label{eq:nogaugeaction}
\end{eqnarray}
where we have defined,
\begin{eqnarray}
\hat{\Phi}_\mu & = &  \{ \hat{A} , \hat{X}_i \}  \nonumber \\
\hat{D}_\mu & = & \{ \partial_\tau + i [ a, \cdot ] , i [ x_i, \cdot ] \} 
\end{eqnarray}
We note that setting the off-diagonal modes to zero yields a trivial massless quadratic theory for the diagonal degrees of freedom. 

\subsection*{Validity of 1-loop integral over off diagonal modes}

Let us for a moment consider taking the diagonal modes, $a(\tau)$ and $x_i(\tau)$ to be constant in time, so that the operators $\hat{D}_\mu$ commute. We also rescale the off diagonal fields to canonically normalize their kinetic term,
\begin{eqnarray}
x^a_i & = & \Lambda \phi^a_i \nonumber \\
( \hat{X}_\mu )_{ab} & = & \frac{\sqrt{\lambda}}{ \Lambda } ( \hat{\Phi}_\mu )_{ab} \; .
\end{eqnarray}
and then we may write the action,
\begin{eqnarray}
S|_{a, x_i const} & = & N  \oint^R d\tau \,
      \tr  \left(  \frac{1}{2} \hat{\Phi}_\mu\, \hat{P}_{\mu\nu} \, \hat{\Phi}_\nu + \frac{\sqrt{\lambda}}{2} [ \hat{\Phi}_\mu , \hat{\Phi}_\nu ]  \hat{D}_\mu \hat{\Phi}_\nu - \frac{\lambda}{4} [ \hat{\Phi}_\mu , \hat{\Phi}_\nu ]^2 \right)
      \label{eq:constaction}
\end{eqnarray}
where we have written down the interaction terms explicitly, and the operator $\hat{P}_{\mu\nu} = \left( \hat{D}_\mu \hat{D}_\nu - \delta_{\mu\nu} (\hat{D}_\alpha)^2 \right)$. Now the Greek index structure of the propagator $\hat{P}_{\mu\nu}$ implies that longitudinal modes (ie. $\Phi_\mu \sim \hat{D}_\mu f$ for some matrix function $f$) have a vanishing propagator. As usual, this vanishing of the longitudinal mode is due to the gauge invariance of the theory. The physical modes, the transverse ones orthogonal to the longitudinal modes, are controlled by the propagator $(\hat{D}_\alpha)^2$. We use the notation that a diagonal matrix $h$ is written in terms of its diagonal entries, $h^a$, so that $h_{ab}= h^a \delta_{ab}$, and then, $\Delta h^{ab} = h^a - h^b$. Then the operator $(\hat{D}_\alpha)^2$ acting on the off diagonal fields is given as, 
\begin{eqnarray}
\left[ (\hat{D}_\alpha)^2 \hat{\Phi}_\mu \right]_{ab} & = & \left( (  \partial_\tau  + i \Delta a^{ab} )^2 - \Lambda^2 | \Delta \phi^{ab} |^2 \right) (\hat{\Phi}_{\mu})_{ab}
\end{eqnarray}
where the left hand side denotes the $(ab)$ entry of the matrix function $(\hat{D}_\alpha)^2 \hat{\Phi}_\mu$ and where $| \Delta \phi^{ab} |^2 = \sum_i (\Delta \phi^{ab}_i)^2$. We see that this operator is diagonal in color, flavour and momentum space, and gives a mass $\Lambda | \Delta \phi^{ab} |$ to the transverse off diagonal mode $(\hat{\Phi}_\mu)_{ab}$. 

We see from the transverse propagator that the physical off-diagonal modes that are most strongly coupled are those that are constant in time. Then as in the main text we observe that a 1-loop integration over the off-diagonal modes, taking $| \Delta \phi^{ab} | \sim O(1)$, is controlled by the same coupling as for the matrix integral, $g = \lambda / ( R \Lambda^4)$. 
Then setting the mass scale $\Lambda$ of the separation of the diagonal components so that,
\begin{eqnarray}
\Lambda & >> & \left( \frac{ \lambda }{ R } \right)^{1/4}
\end{eqnarray}
then we may perform a 1-loop integral over the off diagonal fluctuations. Translating back to the diagonal variables $x^a_i$, we require,
\begin{eqnarray}
R | \Delta x^{ab} | & >> & \left( \lambda R^3 \right)^{1/4}
\end{eqnarray}
so that the off diagonal fields have sufficient mass for the 1-loop approximation to be good.

\subsection*{Performing 1-loop integation over off-diagonal modes}

Now let us return to our action above in equation \eqref{eq:nogaugeaction} now returning the diagonal fields $a, x_i$ to depend on time. We proceed with the 1-loop integration by fixing a constant diagonal gauge for the gauge field so that,
\begin{eqnarray}
a(\tau) = a \; , \qquad \hat{A}(\tau) = 0
\end{eqnarray}
with $a$ being constant in time $\tau$. Since a gauge variation, $\delta A = D_\tau \Lambda$ about this configuration yields,
\begin{eqnarray}
\delta a(\tau) & = & \partial_\tau \lambda \nonumber \\
\delta \hat{A}(\tau) & = & (  \partial_\tau  + i [ a, \cdot ] ) \hat{\Lambda} = \hat{D}_\tau \hat{\Lambda}
\end{eqnarray}
with $\Lambda = \lambda + \hat{\Lambda}$ then the associated Fadeev-Popov determinant is $\det_{\mathrm{diag}}{ \partial_\tau} \det_{\mathrm{off}}{ \hat{D}_\tau }$, where the first determinant is trivial after regulation, and the second is given by,
\begin{eqnarray}
{\det}_{\mathrm{off}}{ \hat{D}_\tau } = \prod_{a \neq b} \det( \partial_\tau + i \Delta a^{ab}) 
\end{eqnarray}
In this gauge, we may write the path integral as,
\begin{eqnarray}
Z & = & \int \left( d a \, d x_i(\tau) \, d {\hat{X}}_i(\tau) \right) \left( {\det}_{\mathrm{off}}{ \hat{D}_\tau } \right) e^{-  S_{g.f.} } \nonumber \\
S_{g.f.} & = & \frac{N}{\lambda}  \oint^R d\tau \,
      \tr  \left(  \frac{1}{2} \dot{x}_i^2 + \frac{1}{2} \hat{X}_i  \left( \hat{D}_i \hat{D}_j - \delta_{ij} (\hat{D}_\tau^2 + \hat{D}_k^2 )\right) \hat{X}_j + O(\hat{X}^4,  \hat{X}^2 \hat{D} \hat{X} ) \right)
\end{eqnarray}
The operator,
$\hat{M}_{ij} = \hat{D}_i \hat{D}_j - \delta_{ij} ( \hat{D}_\tau^2 + \hat{D}_k^2 )$
acting on off-diagonal Hermitian fields has eigenvalue $(\hat{D}_\tau)$ once, and $(\hat{D}_\tau^2 + \hat{D}_k^2)$ eight times. We decompose the diagonal scalars $x_i(\tau)$ into a constant piece, $x_i$,  and Kaluza-Klein modes, $y_i(\tau)$, so,
\begin{eqnarray}
x_i(\tau) = x_i + y_i(\tau) \; , \; \oint d\tau \, y_i(\tau) = 0
\end{eqnarray}
Integrating out the off diagonal modes then yields,
\begin{eqnarray}
Z & = & \int \left( d a \, dx_i \right) d y_i(\tau) e^{- \frac{N}{\lambda}  \oint^R d\tau \,
      \tr  \left(  \frac{1}{2} \dot{y}_i^2  \right) } \left( \prod_{a \neq b} \det( \hat{D}_\tau^2 + \hat{D}_k^2 ) \right)^{-4} 
\end{eqnarray}
where,
\begin{eqnarray}
{\det}_{\mathrm{off}} ( \hat{D}_\tau^2 + \hat{D}_k^2 ) & = & \prod_{a \neq b} \det( \mathcal{O}_{ab}  + \epsilon_{ab} )  \nonumber \\
\mathcal{O}_{ab} & = & - ( \partial_\tau + i \Delta a^{ab} )^2 + | \Delta x^{ab}_i |^2  \nonumber \\
{\epsilon}_{ab} & = &  2 \Delta x^{ab}_i \Delta y^{ab}_i(\tau) + | \Delta y^{ab}_i(\tau) |^2
\end{eqnarray}
and we note that the Fadeev-Popov determinant is exactly cancelled by the eigenvalue $(\hat{D}_\tau)$ of $\hat{M}_{ij}$.  The remaining determinant, $ \det( \hat{D}_\tau^2 + \hat{D}_k^2 )$, represents the effect of integrating over the physical degrees of freedom of the off diagonal fields. We see from the operator $\mathcal{O}$ above that the constant diagonal scalars, $x_i$, give masses $| \Delta x^{ab} |^2$ to the off-diagonal modes, and the non-constant diagonal scalars then interact with these via the $\epsilon$ term above.

As discussed above, taking $R | \Delta x^{ab} | >> \left( \lambda R^3 \right)^{1/4}$ ensures the one loop approximation is valid. However, taking large $| \Delta x^{ab} |^2$ also implies that the operator $\mathcal{O}$ should dominate the operator $\epsilon$ in the 1-loop determinant above. 
We may then proceed by expanding out the determinant above in  powers of the operator $\epsilon$, which then yield an effective potential for the classical moduli $x^a_i$ and corrections to the classical action of the diagonal, non-constant modes,  $y^a_i$. Explicitly,
\begin{eqnarray}
Z  & = & \int \left( d a \, dx_i \right) e^{-V_0[a, x_i] } \int d y_i(\tau) e^{ - S_{tree}[y_i(\tau)]   - S_{1-loop}[a, x_i, y_i(\tau)] }
\end{eqnarray}
where the potential for the constant modes, $V_0$, and the classical and 1-loop interaction terms for  the diagonal non-constant modes, $S_{tree}$ and $S_{1-loop}$ respectively are given as,\begin{eqnarray}
S_{tree}[ y_i(\tau)] & = & \frac{N}{\lambda}  \oint^R d\tau \,  \tr  \left(  \frac{1}{2} \dot{y}_i^2 \right) \nonumber \\
S_{1-loop}[a, x_i, y_i(\tau)] &= & 4 \sum_{a \neq b} \left( \tr( \mathcal{O}_{ab}^{-1} \epsilon_{ab} ) - \frac{1}{2}   \tr( \mathcal{O}_{ab}^{-1} \epsilon_{ab} \mathcal{O}_{ab}^{-1} \epsilon_{ab} ) + O(\epsilon^3) \right) \nonumber \\
V_0[ a, x_i ] &= & 4 \sum_{a \neq b} \log \tr( \mathcal{O}_{ab} ) 
\end{eqnarray}
and we have expanded the 1-loop interaction term to include all quadratic interations on the Kaluza-Klein fields, $y_i(\tau)$. We may evaluate these determinants and traces by expanding in Fourier modes as,
\begin{eqnarray}
y_i(\tau) = \sum_{m=-\infty, m\ne0}^{\infty} y_{i(m)} e^{i (2 \pi / R)  m \tau }
\end{eqnarray}
so that,
\begin{eqnarray}
S_{tree} & = & \frac{2 \pi^2 N}{R \lambda} \sum_a \sum_{n\ne0} n^2 | y^a_{i(n)} |^2
\end{eqnarray}
and the above operators act in the Fourier space as,
\begin{eqnarray}
( \mathcal{O}_{ab} )_{(mn)} &= & \delta_{mn} \left( \left( \frac{2 \pi m}{R} + \Delta a^{ab} \right)^2 + | \Delta x^{ab} |^2 \right)
 \\
( \mathcal{O}^{-1}_{ab} \epsilon_{ab} )_{(mn)}  &= &  \frac{ \delta_{mn} } { \left( \frac{2 \pi m}{R} + \Delta a^{ab} \right)+ | \Delta x^{ab} |^2} \sum_{p\ne0} \left(   2 \Delta x_i^{ab}\Delta y_{i(p)}^{ab} \delta_{m - n - p} + \sum_{q\ne0}  \Delta y_{i(p)}^{ab}\Delta y_{i(q)}^{ab} \delta_{m - n - p - q}\right) \nonumber
\end{eqnarray}
After a suitable regulation (for example Pauli-Villars \cite{Aharony2}) we find,
\begin{eqnarray} 
V_0 & = & 8 \sum_{a<b} \ln \left( \cosh{ R | \Delta  x^{ab} |  } - \cos{ R \Delta A^{ab} } \right)  .
\end{eqnarray}

\subsection*{Validity of 1-loop integation over diagonal non-constant modes}

The condition for 1-loop integration over the off-diagonal modes is $R | \Delta x^{ab} | >> ( \lambda R^3)^{1/4}$. From the above, we see that we may also consider now performing a loop integral over $y^a_{i(n)}$. The most strongly coupled modes are the low momentum modes. For such modes powers of $( \mathcal{O}^{-1} \epsilon )$ contains interaction terms, and for large classical moduli separation $ R | \Delta  x^{ab} |  >> 1$ each power of $y^a_i$ in such an interaction term is suppressed by a factor of 
$\sim 1/ | \Delta  x^{ab} |$. Rescaling the fields $y^a_i$ to obtain a canonical normalization of $S_{tree}$ one then sees that the coupling controlling this 1-loop integration over the $y^a_i$ is,
\begin{eqnarray}
R | \Delta x^{ab} | >> ( \lambda R^3)^{1/2}
\end{eqnarray}
As discussed in the main text, the two conditions $R | \Delta x^{ab} | >> ( \lambda R^3)^{1/4}$ and $R | \Delta x^{ab} | >> ( \lambda R^3)^{1/2}$ are mutually compatible for all dimensionless temperature $t = 1/(\lambda^{1/3} R)$ provided we take sufficiently large $R | \Delta x^{ab} |$.

\subsection*{Performing 1-loop integation over diagonal non-constant modes}

Thus we now perform this 1-loop integral over the diagonal non-constant modes $y^a_i$. Instead of performing it in the maximal range $R | \Delta x^{ab} | >> \max( ( \lambda R^3)^{1/2}, ( \lambda R^3)^{1/4} )$, we instead now focus on the subregion,
 \begin{eqnarray}
 R | \Delta x^{ab} | >> 1
 \end{eqnarray}
 within this range, as this allows some simplifications to be made, and as we are really only interested in the asymptotic behaviour of the remaining potentials at very large $R | \Delta x^{ab} |$.
 Firstly, 
\begin{eqnarray} 
V_0 & \simeq & 8 \sum_{a<b} (R | \Delta  x^{ab} |) 
\end{eqnarray}
Then we may evaluate,
\begin{eqnarray}
\tr( \mathcal{O}_{ab}^{-1} \epsilon_{ab} ) & = &  \sum_{p} \sum_{q\ne0} \frac{  \Delta y_{i(-q)}^{ab}\Delta y_{i(q)}^{ab}} { \left( \frac{2 \pi p}{R} + \Delta a^{ab} \right)+ | \Delta x^{ab} |^2} \nonumber \\
& \simeq & \frac{R}{2 | \Delta x^{ab} |} \sum_{q\ne0} | \Delta y_{i(q)}^{ab} |^2
\end{eqnarray}
for large $R | \Delta x^{ab} | >> 1$, and likewise,
\begin{eqnarray}
\tr( \mathcal{O}_{ab}^{-1} \epsilon_{ab} \mathcal{O}_{ab}^{-1} \epsilon_{ab} ) & = &  \sum_{p} \sum_{q\ne0} \frac{  4 \Delta x_{i}^{ab}\Delta x_{j}^{ab} \Delta y_{i(-q)}^{ab}\Delta y_{j(q)}^{ab}} { \left( \left( \frac{2 \pi p}{R} + \Delta a^{ab} \right)+ | \Delta x^{ab} |^2 \right) \left( \left( \frac{2 \pi (p + n)}{R} + \Delta a^{ab} \right)+ | \Delta x^{ab} |^2 \right)} \nonumber \\
& \simeq & \frac{R}{| \Delta x^{ab} |} \sum_{q\ne0}  \frac{ |\Delta y_{i(q)}^{ab} |^2 }{ | \Delta x^{ab} |^2 + \frac{q^2 \pi^2}{R^2} }
\end{eqnarray}
so that,
\begin{eqnarray}
S_{tree} & = & \frac{2 \pi^2 N}{R \lambda} \sum_a \sum_{n\ne0} n^2 | y^a_{i(n)} |^2 \nonumber \\
S_{1-loop}  & \simeq & 2 R \sum_{a \neq b} \sum_{n\ne 0}   \frac{1}{| \Delta x^{ab} |} ( \Delta{y}^{ab}_{i(n)} )^* \left( \delta_{ij} - \frac{ \Delta x^{ab}_i \Delta x^{ab}_j }{ |\Delta x^{ab}|^2 + \frac{n^2 \pi^2}{R^2}}  \right) (\Delta {y}^{ab}_{j(n)}) + O\left( ( y^a_i)^3 \right) 
\end{eqnarray}
We see that the main contribution to the 1-loop integration over $y^a_i$ is from the classical action $S_{tree}$, but we obtain a small correction from the quadratic terms above in $S_{1-loop}$. We see that the correction is indeed small for $R | \Delta x^{ab} | >> 1$.

In performing this integral, we must take care to recall that $\sum_a y^a_i = 0$ due to the matrix fields $X_i$ being traceless for $SU(N)$ gauge group. 
We split the generator index $a = ( A, N )$ with $A = 1,\ldots,N-1$, and then change to variables,
\begin{eqnarray}
\tilde{y}^A_{i(n)} - \frac{1}{N} \phi_{i(n)} & = & \pi \sqrt{\frac{2}{R \lambda}} y^A_{i(n)} \nonumber \\
- \frac{1}{N} \phi_{i(n)} & = & \pi \sqrt{\frac{2}{R \lambda}} y^N_{i(n)}
\end{eqnarray}
and likewise,
\begin{eqnarray}
\tilde{x}^A_{(n)} - \frac{1}{N} \psi_{i(n)} & = & R x^A_{(n)} \nonumber \\
- \frac{1}{N} \psi_{(n)} & = & R x^N_{(n)}
\end{eqnarray}
with $\phi_{i(n)} = \sum_{A=1}^{N-1} \tilde{y}^A_{i(n)}$ and $\psi_i = \sum_{A=1}^{N-1} \tilde{x}^A_i$. These $(N-1)$ dimensionless variables $\tilde{y}^A_{i(n)}$ now preserve tracelessness of the matrices $X_i$. Changing to these variables gives the quadratic action for $\tilde{y}^A_{i(n)}$ to be,
\begin{eqnarray}
S_{tree} & = & N \sum_{A,B} \sum_{n\ne0} n^2  (\tilde{y}^A_{i(n)})^*  \tilde{y}^B_{i(n)} \left( \delta^{AB} - \frac{1}{N} \right)  \nonumber \\
S_{1-loop} & = & + \frac{ \lambda R^3}{\pi^2} \sum_{A \neq B} \sum_{n\ne 0}   \frac{1}{| \Delta \tilde{x}^{AB} |} ( \Delta{\tilde {y}}^{AB}_{i(n)} )^* \left( \delta_{ij} - \frac{ \Delta \tilde{x}^{AB}_i \Delta \tilde{x}^{AB}_j }{ |\Delta \tilde{x}^{AB}|^2 + n^2 \pi^2}  \right) (\Delta \tilde{y}^{AB}_{j(n)}) \nonumber \\
&  & \quad + \frac{2 \lambda R^3}{\pi^2} \sum_{A} \sum_{n\ne 0}   \frac{1}{| \tilde{x}^{A} |} ({\tilde {y}}^{A}_{i(n)} )^* \left( \delta_{ij} - \frac{ \tilde{x}^{A}_i \tilde{x}^{A}_j }{ | \tilde{x}^{A}|^2 + n^2 \pi^2}  \right) (\tilde{y}^{A}_{j(n)})
\end{eqnarray}
We write the quadratic action $S_{tree} + S_{1-loop}$ as,
\begin{eqnarray}
S_{quad} & = & N \sum_{n\ne0} \sum_{A,B} \sum_{ij} ( \tilde{y}^A_{i(n)} )^* \left( \mathcal{O}'^{AB}_{ij(n)} + \epsilon'^{AB}_{ij(n)} \right) \tilde{y}^B_{j(n)}  \nonumber \\
\mathcal{O}'^{AB}_{ij(n)} & = & n^2 \left( \delta^{AB} - \frac{1}{N} \right) \delta_{ij}   \nonumber \\
\epsilon'^{AB}_{ij(n)} & = & \frac{2 \lambda R^3}{ N \pi^2 } \left( - \frac{1}{| \tilde{x}^{AB} |} \left( \delta_{ij} - \frac{ \tilde{x}^{AB}_i \tilde{x}^{AB}_j }{ | \tilde{x}^{AB}|^2 + n^2 \pi^2}  \right)  \right. \nonumber \\
&& \quad \left. + \delta^{AB} \left[  \frac{1}{| \tilde{x}^{A} |} \left( \delta_{ij} - \frac{ \tilde{x}^{A}_i \tilde{x}^{A}_j }{ | \tilde{x}^{A}|^2 + n^2 \pi^2} \right) + \sum_{C}  \frac{1}{| \tilde{x}^{AC} |} \left( \delta_{ij} - \frac{ \tilde{x}^{AC}_i \tilde{x}^{AC}_j }{ | \tilde{x}^{AC}|^2 + n^2 \pi^2}  \right)  \right] \right) 
\end{eqnarray}
and then performing the 1-loop integral over $\tilde{y}^A_{i(n)}$ and expanding in the correction $\epsilon'$ gives,
\begin{eqnarray}
Z  & \simeq & \int \left( d a \, dx_i \right) e^{-V_0[a, x_i] } (\det \mathcal{O}' )^{-1} e^{ - \tr\left( (\mathcal{O}')^{-1} \epsilon' \right) }
\end{eqnarray}
The determinant $\det( \mathcal{O}' )$ has no moduli dependence and therefore we may neglect it. Computing the trace, we find the result can neatly we written as,
\begin{eqnarray}
\tr\left( (\mathcal{O}')^{-1} \epsilon' \right) = \frac{2 \lambda R^3}{ N \pi^2 }  \sum_{n\ne0} \sum_{a \neq b} \frac{1}{n^2} \frac{1}{R | \Delta x^{ab} |} \left( 9 - \frac{1}{ 1 + \frac{n^2 \pi^2}{R^2 | \Delta x^{ab} |^2} } \right)
\end{eqnarray}
in the original variables $x^a_i$. In the large classical moduli limit, $R | \Delta x^{ab} | >> 1$, this gives,
\begin{eqnarray}
\tr\left( (\mathcal{O}')^{-1} \epsilon' \right) \simeq \frac{32 \lambda R^3}{ 3} \frac{1}{N} \sum_{a < b} \frac{1}{R | \Delta x^{ab} |} 
\end{eqnarray}
and yields the  result given in \eqref{eq:oneloop} of section \ref{sec:BO}.

%
\bibliographystyle{JHEP}
\bibliography{ref}
%

\end{document}